\documentclass[twocolumn,trackchanges]{aastex631}

\usepackage[hyphens]{url}
\usepackage{multirow}

\begin{document}

\title{Physical Parameters and Properties of 20 Cold Brown Dwarfs in JWST}

\correspondingauthor{Shu Wang}
\email{shuwang@nao.cas.cn}

\author[0009-0000-7976-7383]{Zhijun Tu}
\affiliation{CAS Key Laboratory of Optical Astronomy, National Astronomical Observatories, Chinese Academy of Sciences, Beijing 100101, People's Republic of China}
\affiliation{School of Astronomy and Space Sciences, University of Chinese Academy of Sciences, Beijing 100049, People's Republic of China}

\author[0000-0003-4489-9794]{Shu Wang}
\affiliation{CAS Key Laboratory of Optical Astronomy, National Astronomical Observatories, Chinese Academy of Sciences, Beijing 100101, People's Republic of China}

\author{Jifeng Liu}
\affiliation{CAS Key Laboratory of Optical Astronomy, National Astronomical Observatories, Chinese Academy of Sciences, Beijing 100101, People's Republic of China}
\affiliation{School of Astronomy and Space Sciences, University of Chinese Academy of Sciences, Beijing 100049, People's Republic of China}
\affiliation{Institute for Frontiers in Astronomy and Astrophysics, Beijing Normal University, Beijing 102206, People's Republic of China}
\affiliation{New Cornerstone Science Laboratory, National Astronomical Observatories, Chinese Academy of Sciences, Beijing 100012, People's Republic of China}

\begin{abstract}
We present a comprehensive analysis of 20 T and Y dwarfs using spectroscopy from the NIRSpec CLEAR/PRISM and MIRI LRS instruments on the James Webb Space Telescope. To characterize the atmospheric parameters, we utilize two atmospheric model grids: the Sonora Elf Owl and ATMO2020++. The effective temperatures derived from the two models are relatively consistent, and metallicities are both close to solar values. However, significant discrepancies are found in other parameters, particularly in surface gravity, with the values obtained from the Sonora Elf Owl models typically being about 1 dex lower than those from the ATMO2020++ models. Further comparisons using the ATMO2020 models highlight that the adiabatic convective process introduced in the ATMO2020++ models has a significant impact on the determination of surface gravity. Using the fitted effective temperatures and absolute parallaxes from the literature, we derive radii for the brown dwarfs, which range from approximately 0.8 to 1.2\,$R_{\mathrm{Jup}}$. The estimated masses and ages, derived using evolutionary tracks, indicate that most brown dwarfs in our sample have masses below 30\,$M_{\mathrm{Jup}}$ and are younger than 6\,Gyr. Specifically, Y dwarfs have masses ranging from 2 to 20\,$M_{\mathrm{Jup}}$ and ages between 0.1 and 6.7\,Gyr. In addition, We discuss the determination of atmospheric parameters using only NIRSpec or MIRI spectra. Comparisons with results from the combined spectra show that the effective temperatures and surface gravities derived solely from NIRSpec spectra are largely consistent with those obtained from the combined spectra.
\end{abstract}


\section{Introduction} \label{sec:intro}

Brown dwarfs, which occupy the mass range between the heaviest gas giant planets and the lightest stars \citep[$13\,M_{\mathrm{Jup}}\lesssim\mathrm{M}\lesssim73\,M_{\mathrm{Jup}}$,][]{Burrows2001RvMP}, represent a crucial class of objects for understanding the formation and evolution of planetary system, as well as stellar atmospheres. These substellar objects, lacking sufficient mass to sustain stable hydrogen fusion in their cores, display intricate and dynamic atmospheric processes that link planets and stars \citep{Kirkpatrick2005ARA&A}. As such, brown dwarfs provide a unique laboratory for exploring the physical and chemical properties of cool atmospheres, as well as the effect of dust in atmospheres \citep{Allard2001ApJ}.

T and Y dwarfs, the coolest members of the brown dwarf family, are particularly interesting due to their extremely low temperatures and complex atmospheric chemistry. The spectral classification of brown dwarfs spans from the warmer L dwarfs to the cooler T dwarfs, and finally to the even cooler Y dwarfs \citep{Burrows2003ApJ,Burgasser2006ApJ}. T dwarfs, with effective temperatures ranging from approximately 550 to 1300\,K, and Y dwarfs, typically below 500\,K, exhibit prominent molecular absorption features in their spectra. These features include those of water (H$_2$O), methane (CH$_4$), and ammonia (NH$_3$), making these objects rich laboratories for studying the chemistry and physics of low-temperature atmospheres \citep{Cushing2011ApJ,Kirkpatrick2012ApJ}. The investigation of these molecular absorptions provides critical insights into the atmospheric composition and thermal structure.

The advent of the James Webb Space Telescope (JWST) has significantly advanced the study of these cold objects, providing unprecedented sensitivity and resolution in the near-infrared and mid-infrared wavelengths. Instruments such as NIRSpec (Near-Infrared Spectrograph) and MIRI (Mid-Infrared Instrument) on JWST have opened new windows into the detailed characterization of brown dwarf atmospheres, allowing the detection of faint spectral features that were previously inaccessible \citep{NIRSpec,MIRI_LRS}. These advancements enable more precise determinations of the physical and chemical properties of brown dwarfs, such as temperature, pressure, and composition, thus providing deeper insights into their formation and evolutionary histories and refining the atmosphere models \citep{2015ARA&A..53..279M}.

Previous studies have laid the groundwork for our understanding of brown dwarf atmospheres, but the enhanced capabilities of JWST enable us to push the boundaries of this research further. The past research on brown dwarfs has primarily relied on near-infrared ($\sim$0.7--5\,$\mathrm{\mu m}$) spectroscopy. This methodology has facilitated the identification of the main spectral features and determination of the atmospheric compositions of brown dwarfs \citep[e.g.,][]{Cushing2005ApJ,Burgasser2006ApJ_2,Schneider2015ApJ}. Despite its success, this approach has difficulties in providing sufficient precision when analyzing the coldest brown dwarfs due to their extremely low temperatures and faintness in the near-infrared range (especially for less than 3\,$\mathrm{\mu m}$). As a result, new observations, particularly in the mid-infrared, are essential to improve our understanding of these objects. The MIRI instrument on JWST provides the necessary tools to overcome these challenges. With these instruments, we can obtain the spectra across a broader wavelength range, revealing previously inaccessible details about the atmospheric dynamics, composition, and thermal structure of very cold brown dwarfs.

In this study, we present a detailed analysis of 20 T and Y dwarfs observed with NIRSpec and MIRI on JWST. Our sample includes objects spanning a range of spectral types from T6 to $\geq$Y1, providing a comprehensive overview of the atmospheric properties across these classifications. By utilizing the latest models of brown dwarf atmospheres, we aim to derive and compare the physical parameters and properties of these objects. We describe our sample selection standard in Section~\ref{sec:samples} and the data reduction process in Section~\ref{sec:data reduction}. In Section~\ref{sec:modelfit}, we present our methods and the atmospheric models used in our analysis. Section~\ref{sec:Results} focuses on comparing atmospheric parameters derived from the two models and comparing observed spectra with model spectra using the best-fitting parameters. We also examine the variation of spectral features across spectral types and analyze sample properties using evolutionary tracks. In Section~\ref{sec:discuss}, we compare the atmospheric parameters derived from single-instrument spectra (NIRSpec or MIRI) with those obtained by combined spectra, and compare the results of the model fitting to previous work. Finally, we summarize our main results in Section~\ref{sec:conclusion}.

\section{The Sample} \label{sec:samples}

To carry out our analysis, we selected objects identified as brown dwarfs or brown dwarf candidates from the Mikulski Archive for Space Telescopes (MAST)\footnote{\url{https://mast.stsci.edu/portal/Mashup/Clients/Mast/Portal.html}}. The primary selection criterion is that these objects must have both NIRSpec CLEAR/PRISM and MIRI low-resolution spectrometer (LRS) observations. The spectra of NIRSpec CLEAR/PRISM allow for broad wavelength coverage from 0.6 to 5.3\,$\mathrm{\mu m}$ with low spectral resolution ($\text{R}\sim100$), while the spectra of MIRI LRS provide complementary mid-infrared data from 5 to 12\,$\mathrm{\mu m}$ with similar spectral resolution $\text{R}\sim100$.

The screening process was carried out at the end of March 2024. We undertook a thorough effort to screen all publicly available spectra from MAST to ensure they met our selection criteria. This rigorous process resulted in the selection of 20 brown dwarfs and brown dwarf candidates\footnote{The data included in this work can be accessed in MAST via \dataset[DOI]{https://doi.org/10.17909/9mxb-1935}.} Among these, 19 objects were observed as part of the General Observer (GO) program 2302 (PI: M. C. Cushing), while 1 object was observed under the Guaranteed Time Observation (GTO) program 1189 (PI: T. L. Roellig). In Table \ref{tab:samples}, we present a detailed list of the selected objects, including the designations, spectral types, and absolute parallaxes. Hereafter, the full designations of these objects will be abbreviated as HHMM+DD.

Our sample is predominantly composed of T and Y dwarfs. These types are characterized by their cooler temperatures and lower luminosities compared to other brown dwarf classes, making them particularly suitable for the study of low-temperature atmospheric processes. The comprehensive spectral coverage provided by combining NIRSpec and MIRI observations is crucial for a detailed analysis. The capabilities of NIRSpec CLEAR/PRISM in the near-infrared allow for the detection of key molecular features like those caused by H$_2$O, CH$_4$, CO$_2$, and CO, while MIRI's mid-infrared coverage provides access to additional molecular bands such as those caused by H$_2$O, CH$_4$, and NH$_3$ that are vital for constructing accurate atmospheric models.

\begin{deluxetable*}{lccccc}

\tablecaption{Fundamental Properties of Our Sample\label{tab:samples}}
\setlength{\tabcolsep}{0.85 cm}

\tablehead{\colhead{Object Name} & \colhead{Sp. type} & \colhead{Sp. type} & \colhead{$\varpi_{\mathrm{abs}}$} & \colhead{$\varpi_{\mathrm{abs}}$}\\ 
\colhead{} & \colhead{} & \colhead{Ref.} & \colhead{(mas)} & \colhead{Ref.} } 

\startdata
WISE J024714.52+372523.5& T8 & 7 & $64.8\pm2.0$ & 15 \\
WISEPA J031325.96+780744.2& T8.5 & 10 & $ 135.6\pm2.8$ & 2 \\
2MASS J034807.72$-$602227.0& T7 & 1 & $120.1\pm1.8$ & 14 \\
WISE J035934.06$-$540154.6& Y0 & 8 & $ 73.6\pm2.0$ & 2 \\
WISE J043052.92+463331.6& T8 & 7 & $ 96.1\pm2.9$ & 2 \\
WISE J053516.80$-$750024.9& $\geq$Y1: & 8 & $ 68.7\pm2.0$ & 2 \\
WISE J082507.35+280548.5& Y0.5 & 9 & $152.6\pm2.0$ & 2 \\
ULAS J102940.52+093514.6& T8 & 5 & $68.6\pm1.7$ & 15 \\
CWISEP J104756.81+545741.6& Y0 & 3 & $68.1\pm4.9$ & 16 \\
WISE J120604.38+840110.6& Y0 & 9 & $84.7\pm2.1$ & 2 \\
WISEPC J140518.40+553421.4& Y0.5 & 9 & $ 158.2\pm2.6$ & 2 \\
CWISEP J144606.62$-$231717.8& $\geq$Y1 & 3 & $ 103.8\pm5.0$ & 16 \\
WISE J150115.92$-$400418.4& T6 & 6 & $72.8\pm2.3$ & 2 \\
WISEPA J154151.66$-$225025.2& Y1 & 9 & $ 166.9\pm2.0$ & 2 \\
SDSS J162414.37+002915.6& T6 & 1 & $90.9\pm1.2$ & 4 \\
WISEPA J195905.66$-$333833.7& T8 & 10 & $ 83.9\pm2.0$ & 2 \\
WISEPC J205628.90+145953.3& Y0 & 13 & $140.8\pm2.0$ & 2 \\
WISE J210200.15$-$442919.5& T9 & 8 & $92.3\pm1.9$ & 11 \\
WISEA J215949.54$-$480855.2& T9 & 6 & $73.9\pm2.6$ & 2 \\
WISE J220905.73+271143.9& Y0: & 12 & $161.7\pm2.0$ & 2 \\
\enddata

\tablecomments{The fundamental properties of our sample, including Object ID, Spectral Type (Sp. type), and absolute parallax ($\varpi_{\mathrm{abs}}$) along with their respective references. A colon (:) following the spectral type indicates a large uncertainty.}

\tablerefs{(1) \citet{Burgasser2006ApJ}, (2) \citet{Kirkpatrick2021ApJS}, (3) \citet{Meisner2020ApJ}, (4) \citet{Tinney2003AJ}, (5) \citet{Thompson2013PASP}, (6) \citet{Tinney2018ApJS}, (7) \citet{Mace2013ApJS}, (8) \citet{Kirkpatrick2012ApJ}, (9) \citet{Schneider2015ApJ}, (10) \citet{Kirkpatrick2011ApJS}, (11) \citet{Tinney2014ApJ}, (12) \citet{Cushing2014AJ}, (13) \citet{Cushing2011ApJ}, (14) \citet{Kirkpatrick2019ApJS_Parallaxes}, (15) \citet{Best2020AJ}, (16) \citet{Beiler2024arXiv}.}

\end{deluxetable*}

\section{data reduction} \label{sec:data reduction}
\subsection{JWST Pipeline}

The data reduction process for all dwarfs began with the {\sc uncal} files obtained from the Mikulski Archive for Space Telescopes (MAST). We utilized the official JWST pipeline (Version 1.13.4) along with the Calibration Reference Data System (CRDS) version 11.17.14, under the context of 1225.pmap.

In Stage 1 of the pipeline, the {\sc uncal} files, which consist of the raw, non-destructively read ramps, underwent several detector-level corrections. These corrections included superbias subtraction, linearity correction, and dark subtraction. Following these corrections, slopes were fitted to the corrected ramps, resulting in uncalibrated slope images.

Stage 2 involved calibrating the slopes derived in Stage 1. This stage includes various calibration steps common to both NIRSpec slit and MIRI LRS data. Key processes included the assignment of the world-coordinate system, background subtraction, wavelength calibration, and flux calibration. The output of this stage was calibrated slope images, which served as input for Stage 3.

In Stage 3, multiple exposures were combined by aligning them, flagging outlier pixels, and generating a single integrated product, such as extracted 1D spectra or resampled 2D images. 

The official JWST pipeline and CRDS, which have been updated continuously over various versions, currently perform well for most spectra. Consequently, for the spectra of these 20 objects, we uniformly adopted the default parameters provided by the official pipeline for processing. As an example, the differences between the processed spectra of WISE J0359$-$54 in our sample and the spectra of the same object shown by \citet{Beiler2023ApJ} are not significant.

The NIRSpec spectra obtained with the CLEAR/PRISM filter cover an effective wavelength range of 0.6--5.3\,$\mathrm{\mu m}$, while the MIRI LRS spectra span 5--12\,$\mathrm{\mu m}$. The overlapping wavelength range of 5--5.3\,$\mathrm{\mu m}$, in a typical spectrum of a brown dwarf, primarily features $\mathrm{H_2O}$ absorption, with no other significant spectral features. Therefore, cutting and merging spectra at different positions within this wavelength range should not significantly affect the results. In order to test the impact of choosing different cut-off points on parameter fits, using the spectra of WISE J0359$-$54, we have tested a total of 16 cut-off points from 5 to 5.3\,$\mathrm{\mu m}$ at 0.02\,$\mathrm{\mu m}$ intervals to merge the spectra of NIRSpec and MIRI, and fitted the spectra using the methods mentioned in Section~\ref{sec:modelfit}. The fitting results show that parameter fits for spectra with different cut-off points are largely consistent, with differences generally falling within the 1$\sigma$ uncertainty. Additionally, we note that in the MIRI spectra of our sample, there are only four spectral data points between 5 and 5.3\,$\mathrm{\mu m}$, with the third data point always having a wavelength greater than 5.19\,$\mathrm{\mu m}$. Therefore, considering the higher resolution of the NIRSpec spectra within the overlapping wavelength range, but with the relatively large flux errors at the reddest end, and the necessity of maintaining consistency in the use of MIRI spectra for fitting the wavelength correction parameters, we selected the middle data point of MIRI spectrum---5.19\,$\mathrm{\mu m}$ as the cut-off point to merge the NIRSpec and MIRI spectra.

Additionally, the fluxes blueward of 0.95\,$\mathrm{\mu m}$ were very weak and exhibited large flux uncertainties, with no significant spectral features present. Thus, this portion of the spectrum was excluded from our analysis. The final reduced spectra for each object cover the wavelength range of 0.95--12\,$\mathrm{\mu m}$. This standardized approach ensures consistency and reliability of the spectral data used for subsequent analysis.

\subsection{Wavelength and Flux Calibration} \label{subsec:LRS CAL}

After processing with the official JWST pipeline, the wavelength calibration and absolute flux calibration of the MIRI LRS still exhibit non-negligible uncertainties. The wavelength calibration of MIRI LRS is accurate to within $\pm20$\,nm across most wavelengths, but slightly larger discrepancies exist at the shortest wavelengths\footnote{See details in \url{https://jwst-docs.stsci.edu/known-issues-with-jwst-data/miri-known-issues/miri-lrs-known-issues}}. We will discuss the details of such wavelength correction in Section~\ref{subsec:method}, as it is part of the parameters in the spectral fitting.

The goal for absolute flux calibration of JWST spectra was $\sim$10--15\% \citep{Gordon2022AJ}, while the goal for photometry of JWST was to reach $\sim$5\%. As reported by \citet{Beiler2024arXiv}, they generated synthetic WISE (W1 and W2) and Spitzer ([3.6] and [4.5]) photometry from the NIRSpec spectra of 23 brown dwarfs. Comparison with the generated photometry from the NIRSpec spectra and observed photometry reveals that the differences between generated photometry and observed photometry of W1, W2, and [4.5] are all evenly distributed around zero, except for [3.6]. The generated [3.6] photometry systematically shows $\sim$0.3 magnitudes fainter than the observed photometry, which may be attributed to a light leak or a small discrepancy in the red edge of the [3.6] transmission curve. Overall, the current NIRSpec spectra are well matched to most photometry, so we do not make further absolute flux calibration for the NIRspec spectra.

For the absolute flux calibration of MIRI LRS spectra, \citet{Beiler2024arXiv} reported a 4\% offset between the generated photometry from MIRI spectra and F1000W photometry. Therefore, we use the MIRI F1000W photometry to calibrate the flux of MIRI spectra. For 19 objects with F1000W photometry, we calculate the flux densities $f_{\mathrm{\nu,F1000W}}$ from the F1000W photometry, as well as the average flux densities $f_{\lambda,\mathrm{spec}}$ of the spectra under the F1000W filter using the equation:
\begin{equation}
\left\langle f_{\lambda,\mathrm{spec}} \right\rangle = \frac{\int f_{\lambda} R_\lambda \lambda d\lambda}{\int R_\lambda \lambda d\lambda},
\end{equation}
where $f_\lambda$ is the observed flux density of each object and $R_\lambda$ is the bandpass function of the F1000W filter, which is taken from the Spanish Virtual Observatory (SVO)\footnote{\url{http://svo2.cab.inta-csic.es/svo/theory/fps3/index.php}}. The $f_{\lambda,\mathrm{spec}}$ can be converted to $f_{\nu,\mathrm{spec}}$ using the equation:
\begin{equation}
\left\langle f_{\nu,\mathrm{spec}} \right\rangle = \frac{\lambda_\mathrm{pivot}^2}{c} \left\langle f_{\lambda,\mathrm{spec}} \right\rangle,
\end{equation}
where $\lambda_\mathrm{pivot}$ is the pivot wavelength of the F1000W filter ($\lambda_\mathrm{pivot}=9.954\,\mathrm{\mu m}$), and $c$ is the speed of light. Therefore, the scaling factor for the absolute flux calibration of the spectrum is defined as $R=f_{\nu,\mathrm{F1000W}}/f_{\nu,\mathrm{spec}}$.

For the F1000W photometry, we used two methods to derive $f_{\mathrm{\nu,F1000W}}$. The first method involved processing the photometry with the same version of the JWST pipeline and CRDS, which utilizes the \textsc{photutils} Python package for photometry. The second method employed the \textsc{DOLPHOT 2.0} software \citep{dolphot1,dolphot2}, following the guidelines of the JWST/MIRI module provided by \citet{dolphot2}. We yielded a scaling factor from each method. The MIRI spectra of the objects with F1000W photometry are calibrated with the average scaling factor of the two methods. We also calculate the average scaling factor for all objects with F1000W photometry. The average factor obtained by the JWST pipeline was $R_1=1.038$, and the average factor obtained by \textsc{DOLPHOT 2.0} was $R_2=1.040$. These scaling factors are consistent with that reported by \citet{Beiler2024arXiv}. For one object, 2MASS J0348$-$60, without F1000W photometry, we adopt the average factor of 1.04 to calibrate the spectral flux of MIRI.

By implementing this refined calibration process, we achieve a higher accuracy in absolute flux calibration, which is essential to ensure consistency between photometry and spectra, and for making precise comparisons with theoretical models.

\section{Model fitting} \label{sec:modelfit}
\subsection{Models} \label{subsec: model}

Self-consistent radiative-convective equilibrium (RCE) atmospheric models are widely used to study the atmospheres of brown dwarfs and giant planets. The fundamental properties of brown dwarfs are often derived by fitting synthetic spectra generated from grids of atmospheric models. However, the extremely low temperature leads to a chemical equilibrium that favors the formation of molecules. The complex chemistry arising from molecule-rich atmospheres complicates the determination of the fundamental properties of brown dwarfs from their spectra \citep{Marley2015ARA&A,Phillips2020A&A}.

Consequently, multiple theoretical atmospheric models for brown dwarfs exist, yet none have achieved a perfect match with observations. For our analysis, we adopted two different grids of models: ATMO2020++ \citep{Meisner2023AJ} and Sonora Elf Owl \citep{Mukherjee2024ApJ}. Both models are RCE cloudless atmospheric models incorporating disequilibrium chemistry. Disequilibrium chemistry due to vertical mixing is a common feature in brown dwarf observations \citep{Leggett2015ApJ,Leggett2017ApJ}. Additionally, cloudless models have been verified to be more consistent with the atmospheres of T and Y dwarfs \citep{Allard2001ApJ,Tremblin2015ApJ}. The primary differences between these two models lie in their opacity databases, parameter assumptions, and treatment of convective processes.

ATMO2020++ is a modified version of the ATMO2020 models \citep{Phillips2020A&A}, in which the temperature gradient was reduced due to the convective instabilities triggered by chemical transitions such as CO/CH$_4$ and N$_2$/NH$_3$ transitions. This modification of the model is implemented by introducing the adiabatic convective processes and the corresponding effective adiabatic index. The effective adiabatic index is set at a constant value of 1.25 (See details in \citealt{Meisner2023AJ}). Disequilibrium chemistry is used with a vertical eddy diffusion coefficient, $K_\mathrm{zz}$, which is not treated as a free parameter. Instead, $K_\mathrm{zz}$ is determined by the relationship $\log K_\mathrm{zz}=15-2\log g$. The grids of this model include the following free parameters: effective temperatures ($T_\mathrm{eff}$), surface gravity ($\log g$), and atmospheric metallicity ($[\mathrm{M/H}]$). The ranges of these parameters are present in Table~\ref{tab:para_range}. The increments of $T_\mathrm{eff}$ is 25\,K between 250 and 300\,K, 50\,K between 300 and 500 K, and 100\,K between 500 and 1200\,K. The increments of $\log g$ is 0.5 dex.

The Sonora Elf Owl models represent the latest generation in the Sonora model series, following the Sonora Bobcat \citep{Marley2021ApJ} and Sonora Cholla \citep{Karalidi2021ApJ} models. Compared to its predecessors, the Sonora Elf Owl model introduces more free parameters and covers wider parameter spaces. It should be noted that the PH$_3$ abundance is treated separately from the general disequilibrium scheme, but instead set to a chemical equilibrium abundance, due to the non-detection of PH$_3$ in most brown dwarf atmospheres. The model grids include five parameters: $T_\mathrm{eff}$, $\log g$, $[\mathrm{M/H}]$, carbon-to-oxygen ratio ($\mathrm{C/O}$), and vertical eddy diffusion coefficient ($\log K_\mathrm{zz}$). The ranges of these parameters are shown in Table~\ref{tab:para_range}. The increments of $T_\mathrm{eff}$ and $\log g$ can be found in \citet[][Table~2]{Mukherjee2024ApJ}.

\begin{deluxetable}{ccc}
\tablecaption{Atmospheric Model Parameter Ranges\label{tab:para_range}}

\tablehead{\colhead{Parameter Name} & \colhead{Sonora Elf Owl} & \colhead{ATMO2020++}} 
\startdata
$T_\mathrm{eff}$(K) &  275--2400 &    250--1200 \\
$\log g\,[\mathrm{cm\,s^{-2}}]$ &  3.25--5.5 &    2.5--5.5 \\
$\mathrm{[M/H]}$ &  $-$1.0, $-$0.5, +0.0, &    $-$1.0, $-$0.5, +0.0 \\
 &+0.5, +0.7, +1.0 & \\
$\log K_\mathrm{zz}\,[\mathrm{cm^2\,s^{-1}}]$ &  2, 4, 7, 8, 9 &   $15-2\log g$ \\
C/O$^a$  & 0.22, 0.458, 0.687, 1.12  &  * \\
\enddata
\tablecomments{The solar carbon–to–oxygen ratio is 0.458.}
\end{deluxetable}

\begin{deluxetable*}{llllrlllll}[htp!]

\tabletypesize{\footnotesize}

\tablecaption{The Fitted Parameters Derived with Sonora Elf Owl Models\label{tab:params_elf}}

\tablehead{\colhead{Object ID} & \colhead{$\log K_\mathrm{zz}$} & \colhead{$T_\mathrm{eff}$} & \colhead{$\log g$} & \colhead{[M/H]} & \colhead{C/O} & \colhead{$A_\lambda$} & \colhead{$C_\lambda$} & \colhead{$\log(R^2/D^2)$} \\ 
\colhead{} & \colhead{[$\mathrm{cm^2\,s^{-1}}$]} & \colhead{(K)} & \colhead{[$\mathrm{cm\,s^{-2}}$]} & \colhead{} & \colhead{} & \colhead{($\times 10^{-3}$)} & \colhead{($\times 10^{-3}$\,$\mathrm{\mu m}$)} & \colhead{} } 

\startdata
WISE J0247+37&$2.27\pm 0.03$&$658.73^{+0.5}_{-0.52}$&$4.01\pm 0.01$&$-0.13\pm 0.01$&$0.38\pm 0.01$&$-4.83^{+0.04}_{-0.14}$&$47.79^{+0.83}_{-0.97}$&$-19.75\pm 0.01$\\
WISEPA J0313+78&$2.74\pm 0.01$&$572.74^{+0.28}_{-0.29}$&$3.25\pm 0.01$&$-0.08\pm 0.01$&$0.22\pm 0.01$&$-3.91\pm 0.02$&$39.01^{+0.14}_{-0.15}$&$-19.00\pm 0.01$\\
2MASS J0348$-$60&$2.00\pm 0.01$&$883.32\pm 0.08$&$3.54\pm 0.01$&$-0.23\pm 0.01$&$0.25\pm 0.01$&$-6.27\pm 0.01$&$58.19\pm 0.02$&$-19.21\pm 0.01$\\
WISE J0359$-$54&$5.30\pm 0.02$&$457.26^{+0.35}_{-0.38}$&$3.25\pm 0.01$&$-0.19\pm 0.01$&$0.22\pm 0.01$&$-4.22\pm 0.01$&$41.03^{+0.99}_{-0.04}$&$-19.51\pm 0.01$\\
WISE J0430+46&$5.10\pm 0.03$&$539.21^{+0.54}_{-0.52}$&$3.73\pm 0.01$&$-0.56\pm 0.01$&$0.22\pm 0.01$&$-8.59^{+0.01}_{-0.02}$&$83.88^{+0.19}_{-0.09}$&$-19.46\pm 0.01$\\
WISE J0535$-$75&$4.46\pm 0.01$&$399.00^{+0.34}_{-0.33}$&$3.25\pm 0.01$&$-0.05\pm 0.01$&$0.22\pm 0.01$&$-5.19\pm 0.01$&$49.59^{+0.02}_{-0.01}$&$-19.16\pm 0.01$\\
WISE J0825+28&$5.19\pm 0.01$&$387.18^{+0.21}_{-0.2}$&$3.25\pm 0.01$&$0.25\pm 0.01$&$0.22\pm 0.01$&$-5.32\pm 0.01$&$52.61^{+0.05}_{-0.03}$&$-18.81\pm 0.01$\\
ULAS J1029+09&$2.08\pm 0.03$&$769.28^{+0.36}_{-0.37}$&$3.25\pm 0.01$&$0.20\pm 0.01$&$0.25\pm 0.01$&$-3.00\pm 0.01$&$30.23^{+0.02}_{-0.03}$&$-19.75\pm 0.01$\\
CWISEP J1047+54&$4.75\pm 0.02$&$427.57^{+0.24}_{-0.25}$&$3.67\pm 0.01$&$0.31\pm 0.01$&$0.33\pm 0.01$&$-3.15\pm 0.01$&$30.10^{+0.04}_{-0.03}$&$-19.68\pm 0.01$\\
WISE J1206+84&$9.00\pm 0.01$&$439.22\pm 0.23$&$3.25\pm 0.01$&$0.22\pm 0.01$&$0.50\pm 0.01$&$-4.28\pm 0.01$&$40.26^{+0.06}_{-0.05}$&$-19.27\pm 0.01$\\
WISEPC J1405+55&$4.28\pm 0.01$&$388.65^{+0.28}_{-0.29}$&$3.25\pm 0.01$&$-0.04\pm 0.01$&$0.22\pm 0.01$&$-4.21\pm 0.01$&$39.98^{+0.03}_{-0.02}$&$-18.77\pm 0.01$\\
CWISEP J1446$-$23&$4.90\pm 0.02$&$362.19^{+0.26}_{-0.25}$&$3.25\pm 0.01$&$0.38\pm 0.01$&$0.25\pm 0.01$&$-3.11\pm 0.01$&$30.80\pm 0.02$&$-19.19\pm 0.01$\\
WISE J1501$-$40&$3.76^{+0.03}_{-0.02}$&$993.52^{+0.48}_{-0.49}$&$4.75\pm 0.01$&$0.30\pm 0.01$&$0.60\pm 0.01$&$-3.93^{+0.04}_{-0.05}$&$40.05^{+0.76}_{-0.35}$&$-19.88\pm 0.01$\\
WISEPA J1541$-$22&$5.57\pm 0.02$&$407.47^{+0.2}_{-0.19}$&$3.25\pm 0.01$&$0.28\pm 0.01$&$0.22\pm 0.01$&$-5.32\pm 0.01$&$53.72^{+0.02}_{-0.03}$&$-18.73\pm 0.01$\\
SDSS J1624+00&$2.04\pm 0.01$&$1026.58\pm 0.3$&$4.46\pm 0.01$&$-0.14\pm 0.01$&$0.39\pm 0.01$&$-3.34^{+0.2}_{-0.43}$&$34.08^{+3.78}_{-1.42}$&$-19.58\pm 0.01$\\
WISEPA J1959$-$33&$2.71\pm 0.02$&$781.23^{+0.33}_{-0.35}$&$3.25\pm 0.01$&$0.20\pm 0.01$&$0.27\pm 0.01$&$-3.58\pm 0.01$&$30.02^{+0.03}_{-0.02}$&$-19.48\pm 0.01$\\
WISEPC J2056+14&$5.24^{+0.01}_{-0.02}$&$471.71^{+0.31}_{-0.3}$&$3.25\pm 0.01$&$0.22\pm 0.01$&$0.25\pm 0.01$&$-4.19\pm 0.01$&$44.35\pm 0.05$&$-18.87\pm 0.01$\\
WISE J2102$-$44&$2.77^{+0.02}_{-0.03}$&$573.85^{+0.91}_{-0.62}$&$3.56\pm 0.01$&$-0.05\pm 0.01$&$0.30\pm 0.01$&$-3.88\pm 0.01$&$41.62^{+0.05}_{-0.07}$&$-19.40\pm 0.01$\\
WISEA J2159$-$48&$2.78\pm 0.04$&$549.49^{+0.9}_{-0.75}$&$3.92\pm 0.01$&$-0.16\pm 0.01$&$0.25\pm 0.01$&$-5.22\pm 0.01$&$50.96^{+0.02}_{-0.03}$&$-19.59\pm 0.01$\\
WISE J2209+27&$5.22\pm 0.02$&$362.42\pm 0.23$&$3.25\pm 0.01$&$0.22\pm 0.01$&$0.22\pm 0.01$&$-4.79\pm 0.01$&$47.37^{+0.05}_{-0.03}$&$-18.77\pm 0.01$\\
\enddata

\tablecomments{The parameters and uncertainties are taken from the median values and 68\% probability interval of the marginalized posterior distributions, delineated by the 16th, 50th, and 84th quantiles. The uncertainties less than 0.01 are approximated to 0.01 in this table. We note that these fits only include the observed flux errors. The estimates of uncertainties caused by linear interpolation are presented in Appendix~\ref{app:Linear Interpolation Accuracy}.}

\end{deluxetable*}
\subsection{Fitting Methods} \label{subsec:method}

In our analysis, we utilized all available grid points from both models to ensure comprehensive coverage of all parameters. The spectral resolutions of the Sonora Elf Owl model spectra are remarkably high. To balance computational efficiency and the accuracy required for fitting wavelength correction parameters, we resampled the Sonora Elf Owl model spectra to spectra with a wavelength interval of 0.001\,$\mathrm{\mu m}$. This resampling allowed us to reduce the data size while maintaining sufficient detail. For ATMO2020++ models, We retained the original spectra due to the relatively low spectral resolution ($R\sim3000$). Subsequently, we performed linear interpolation across the model grids for all parameters on the model spectra. The linear interpolation used the \textsc{RegularGridInterpolator} function based on the Python \textsc{Scipy} module. Further discussion on the reliability of the linear interpolation is provided in Appendix~\ref{app:Linear Interpolation Accuracy}.

Since the resolutions of the resampled spectra of the Sonora Elf Owl models and spectra of the ATMO2020++ models did not match that of the observed spectra, we applied convolution to the linearly interpolated model spectra. We used the approximate resolving power expressions provided by \citet{Beiler2023ApJ} (Equations 3 and 4) for NIRSpec CLEAR/PRISM and MIRI LRS, respectively. The resulting wavelength-resolving power relationships for NIRSpec CLEAR/PRISM and MIRI LRS were generally consistent with those reported by \citet{NIRSpec} and \citet{MIRI_LRS}.

To correct the observed spectral wavelength of the MIRI LRS, we applied a linear correction using the equation:
\begin{equation} \label{eq:wave_cor}
\lambda_\mathrm{cor}= \lambda_\mathrm{MIRI}+\Delta\lambda=(1+A_\lambda)\lambda_\mathrm{MIRI}+C_\lambda,
\end{equation}
where $\lambda_\mathrm{cor}$ is the corrected wavelength and $\lambda_\mathrm{MIRI}$ is the observed wavelength of the MIRI LRS spectra. The parameters $A_\lambda$ and $C_\lambda$ are included in our fitting process as free parameters and are applied during convolution, as they determined which segment of the model spectra corresponded to specific wavelengths in the observed spectra.

To derive posterior probability distributions and Bayesian evidence, we utilized the nested sampling Monte Carlo algorithm "MLFriends" \citep{untranest2016,ultranest2019} with the open-source code \textsc{UltraNest} \citep{ultranest}. The nested sampling algorithm is particularly well-suited for this application due to its efficiency in exploring complex, multi-dimensional parameter spaces and accurately estimating the evidence. The likelihood function used in our fitting procedure is expressed as
\begin{equation}
\ln L = -0.5 \sum_i \left[ \frac{(F_i-\frac{R^2}{D^2} M_i)^2}{\sigma^2_i} + \ln(2\pi\sigma^2_i) \right],
\end{equation}
where $F$ is the observed flux density, and $M$ is the flux density of model spectra after convolution. The $\sigma$ represents the flux errors of the observed spectra. The term $\frac{R^2}{D^2}$, treated as a free parameter, is a scaling factor for the model spectra, where $R$ is the radius of the brown dwarf and $D$ is its distance.

In total, we have eight parameters to fit with the Sonora Elf Owl models and six parameters with the ATMO2020++ models, including the model grid parameters, wavelength correction parameters, and logarithm of the scaling factor $\log(R^2/D^2)$.

The prior distributions for the model parameters are set as uniform distributions over the entire parameter spaces of the models (refer to section \ref{subsec: model}). For $\log(R^2/D^2)$ parameters, it is uniformly distributed between $-30$ and $-10$. To estimate the prior distributions for the wavelength correction parameters, we performed preliminary fits based on nested sampling using broad prior distributions for the wavelength correction parameters. The resulting fits showed that, for most sources, $A_\lambda$ values fell between $-0.009$ and $-0.004$, and $C_\lambda$ ranged from 0.04 to 0.09\,$\mathrm{\mu m}$. These results indicate that the observed MIRI spectra are slightly blueshifted relative to the model spectra. However, for some sources, $A_\lambda$ were positive, and $C_\lambda$ were negative, which implies that the observed MIRI spectra are redshifted. Upon inspecting the observed spectra for these sources with anomalous wavelength correction parameters, we determined that the true values of $A_\lambda$ and $C_\lambda$ should also follow the trend seen in the majority of sources, where $A_\lambda$ is negative and $C_\lambda$ is positive.

In order to obtain more appropriate prior distributions for the wavelength correction parameters, we used the flux peaks at 6.3, 7.6, and 10.2\,$\mathrm{\mu m}$, and compared the wavelength deviations of these peaks between observed spectra and best-fitting model spectra\footnote{The parameters of the best-fitting model are derived through nested sampling with broad prior distributions for the wavelength correction parameters, i.e., the preliminary fits.} across all objects and used Equation~(\ref{eq:wave_cor}) to fit the $A_\lambda$ and $C_\lambda$. For the Sonora Elf Owl model spectra, we derived the average $A_\lambda=-0.011$ and $C_\lambda=0.106$\,$\mathrm{\mu m}$. And for ATMO2020++ models, the average $A_\lambda=-0.009$ and $C_\lambda=0.080$\,$\mathrm{\mu m}$. For all the wavelength correction parameters in the two models, the $A_\lambda$ values range from $-$0.017 to $-$0.006, while the $C_\lambda$ values are within 0.059 and 0.146\,$\mathrm{\mu m}$. The minimum $A_\lambda$ is associated with the largest $C_\lambda$ and vice versa. And the ratio of $C_\lambda$ to $A_\lambda$ for each object is always around $-$10\,$\mathrm{\mu m}$.  Compared to wavelength correction parameters derived by the preliminary fits through nested sampling, the manual fits of the three peaks yield a smaller $A_\lambda$ and a larger $C_\lambda$, indicating a larger wavelength difference between the observed and model spectra around 5\,$\mathrm{\mu m}$. Considering the results obtained by the two methods, we set the uniform prior distributions for $A_\lambda$ between $-0.020$ and $-0.003$, and for $C_\lambda$ between $0.03$\,$\mathrm{\mu m}$ and $0.2$\,$\mathrm{\mu m}$.

\begin{deluxetable*}{lllrllll}

\tablecaption{The Fitted Parameters Derived with ATMO2020++ Models\label{tab:params_atmo}}
\setlength{\tabcolsep}{0.45cm}

\tablehead{\colhead{Object ID} & \colhead{$T_\mathrm{eff}$} & \colhead{$\log g$} & \colhead{[M/H]}  & \colhead{$A_\lambda$} & \colhead{$C_\lambda$} & \colhead{$\log(R^2/D^2)$} \\ 
\colhead{}  & \colhead{(K)} & \colhead{[$\mathrm{cm\,s^{-2}}$]} &  \colhead{} & \colhead{($\times 10^{-3}$)} & \colhead{($\times 10^{-3}$\,$\mathrm{\mu m}$)} & \colhead{}} 

\startdata
WISE J0247+37&$660.34^{+0.53}_{-0.52}$&$5.11\pm 0.01$&$0.00\pm 0.01$&$-7.83\pm 0.01$&$73.65^{+0.09}_{-0.08}$&$-19.84\pm 0.01$\\
WISEPA J0313+78&$564.47^{+0.36}_{-0.35}$&$4.77\pm 0.01$&$0.00\pm 0.01$&$-6.43\pm 0.01$&$59.82\pm 0.01$&$-19.06\pm 0.01$\\
2MASS J0348$-$60&$811.76\pm 0.06$&$5.24\pm 0.01$&$0.00\pm 0.01$&$-8.06\pm 0.01$&$75.88\pm 0.04$&$-19.20\pm 0.01$\\
WISE J0359$-$54&$475.36^{+0.35}_{-0.37}$&$4.57\pm 0.01$&$-0.10\pm 0.01$&$-5.41\pm 0.01$&$49.50^{+0.05}_{-0.1}$&$-19.62\pm 0.01$\\
WISE J0430+46&$606.01\pm 0.3$&$5.50\pm 0.01$&$-0.06\pm 0.01$&$-10.99\pm 0.01$&$104.54^{+0.08}_{-0.01}$&$-19.66\pm 0.01$\\
WISE J0535$-$75&$409.72^{+0.16}_{-0.15}$&$4.58\pm 0.01$&$0.00\pm 0.01$&$-3.62\pm 0.01$&$30.02\pm 0.02$&$-19.26\pm 0.01$\\
WISE J0825+28&$374.78\pm 0.2$&$3.79\pm 0.01$&$-0.19\pm 0.01$&$-3.69\pm 0.01$&$30.85^{+0.07}_{-0.08}$&$-18.83\pm 0.01$\\
ULAS J1029+09&$739.56^{+0.38}_{-0.37}$&$4.79\pm 0.01$&$0.00\pm 0.01$&$-7.66\pm 0.01$&$73.59^{+0.05}_{-0.07}$&$-19.76\pm 0.01$\\
CWISEP J1047+54&$381.01^{+0.24}_{-0.23}$&$2.50\pm 0.01$&$-0.98\pm 0.01$&$-3.79\pm 0.01$&$32.86\pm 0.01$&$-19.64\pm 0.01$\\
WISE J1206+84&$466.42^{+0.2}_{-0.21}$&$4.08\pm 0.01$&$0.00\pm 0.01$&$-3.69\pm 0.01$&$30.07\pm 0.02$&$-19.38\pm 0.01$\\
WISEPC J1405+55&$402.37\pm 0.13$&$4.58\pm 0.01$&$0.00\pm 0.01$&$-3.71\pm 0.01$&$30.00\pm 0.01$&$-18.88\pm 0.01$\\
CWISEP J1446$-$23&$362.77^{+0.13}_{-0.14}$&$3.88\pm 0.01$&$0.00\pm 0.01$&$-3.63\pm 0.01$&$30.38^{+0.09}_{-0.06}$&$-19.27\pm 0.01$\\
WISE J1501$-$40&$1016.20^{+0.41}_{-0.4}$&$5.43\pm 0.01$&$0.00\pm 0.01$&$-8.08^{+0.03}_{-0.01}$&$76.08^{+0.1}_{-0.21}$&$-19.95\pm 0.01$\\
WISEPA J1541$-$22&$394.63^{+0.29}_{-0.31}$&$3.69\pm 0.01$&$-0.21\pm 0.01$&$-3.39\pm 0.01$&$30.01^{+0.02}_{-0.01}$&$-18.75\pm 0.01$\\
SDSS J1624+00&$1034.36^{+0.3}_{-0.29}$&$5.50\pm 0.01$&$0.00\pm 0.01$&$-8.31^{+0.04}_{-0.92}$&$77.36^{+9.25}_{-0.28}$&$-19.63\pm 0.01$\\
WISEPA J1959$-$33&$766.56^{+0.39}_{-0.38}$&$4.73\pm 0.01$&$0.00\pm 0.01$&$-4.25\pm 0.01$&$39.84^{+0.08}_{-0.06}$&$-19.52\pm 0.01$\\
WISEPC J2056+14&$458.55^{+0.13}_{-0.14}$&$4.14\pm 0.01$&$0.00\pm 0.01$&$-4.77\pm 0.01$&$44.91\pm 0.01$&$-18.89\pm 0.01$\\
WISE J2102$-$44&$545.85^{+0.38}_{-0.39}$&$4.75\pm 0.01$&$0.00\pm 0.01$&$-5.06^{+0.02}_{-0.01}$&$49.39^{+0.06}_{-0.12}$&$-19.42\pm 0.01$\\
WISEA J2159$-$48&$541.46^{+0.45}_{-0.43}$&$5.20\pm 0.01$&$0.00\pm 0.01$&$-7.59\pm 0.01$&$72.80\pm 0.02$&$-19.65\pm 0.01$\\
WISE J2209+27&$362.93^{+0.1}_{-0.11}$&$4.00\pm 0.01$&$0.00\pm 0.01$&$-3.60\pm 0.01$&$31.48^{+0.02}_{-0.01}$&$-18.85\pm 0.01$\\
\enddata

\tablecomments{Similar to Table \ref{tab:params_elf}, but using the ATMO2020++ model.}

\end{deluxetable*}

\section{Results} \label{sec:Results}

\subsection{The model-fitting results} \label{subsec:model_fit_results}

\begin{figure}[htp!]
\centering 
\includegraphics[width=0.45\textwidth]{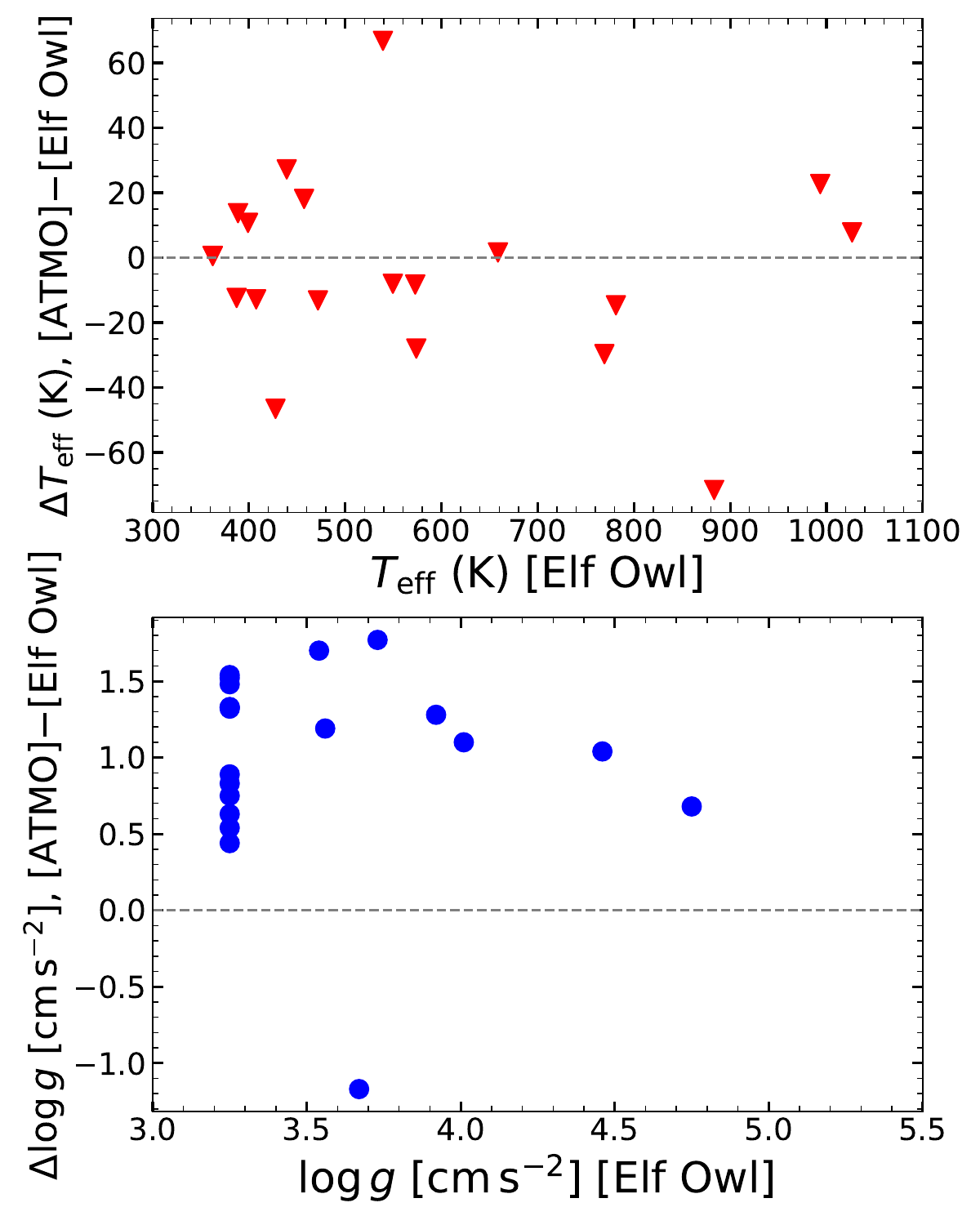}
\caption{Comparisons of effective temperature and surface gravity derived from the two models. Parameter values on the x-axis are derived from the Sonora Elf Owl models, while those on the y-axis represent the difference between the ATMO2020++ and Sonora Elf Owl models.\\\label{fig:teff_log}} 
\end{figure}

The fitting results from the Sonora Elf Owl and ATMO2020++ models are detailed in Tables \ref{tab:params_elf} and \ref{tab:params_atmo}. The model-fit comparisons and corner plots for all sources are publicly available online\footnote{\url{https://github.com/LuShenJ/Parameters_of_20_Cold_Brown_Dwarfs_in_JWST}}. The fitting parameters in both tables show that the uncertainties of the model parameters are significantly smaller than the intervals of the model grids, revealing the underestimation of these uncertainties. In our analysis, only the uncertainties of the observed flux densities were considered, not including uncertainties from the models themselves or those introduced by our methods, such as the uncertainties associated with linear interpolation. More discussion on the linear interpolation errors can be found in Appendix~\ref{app:Linear Interpolation Accuracy}.

Figure~\ref{fig:teff_log} shows a comparison of the main parameters, effective temperature (top) and surface gravity (bottom). Both models yield consistent effective temperatures. However, there is a notable discrepancy in surface gravity between the two models. Specifically, the $\log g$ values derived from the ATMO2020++ models are systematically higher, by approximately 1 dex on average, compared to those obtained by the Sonora Elf Owl models. Determining surface gravity by spectroscopy is generally more challenging than determining the effective temperature and is often associated with larger uncertainties. In particular, 12 objects using the Sonora Elf Owl models converge to the boundary value $\log g = 3.25$\,[$\mathrm{cm\,s^{-2}}$], with a total of 17 objects exhibiting $\log g$ values less than 4.

The metallicity parameter space of the ATMO2020++ models is restricted to sub-solar metallicity, whereas the Sonora Elf Owl models hold both sub-solar and super-solar metallicities. This difference, along with the distinct opacity databases utilized by the two models, results in discrepancies in the fitted [M/H] values derived using the two models. In the analysis using the Sonora Elf Owl models, the metallicities of the 20 objects are evenly distributed, with 10 objects showing super-solar metallicities and the other 10 objects showing sub-solar metallicities. All objects, except WISE J0430+46 with $\text{[M/H]} = -0.56$, have metallicities between $-$0.3 and 0.4, which is close to the solar value. On the other hand, the results of ATMO2020++ models show that 15 out of 20 objects converge to the solar metallicity of 0, with 19 objects having metallicities greater than $-$0.3. The only exception is CWISEP J1047+54, which has a metallicity of $-$0.98. All of the objects with measured parallaxes are located within 20\,pc of the Sun, suggesting that they are expected to have solar-like metallicities \citep{Leggett2017ApJ}.

The fitted C/O ratios, a parameter in the Sonora Elf Owl models, indicate that nearly all objects possess C/O ratios lower than the solar value of 0.458 \citep{Lodders2009LanB}. Only WISE J1206+84 and WISE J1501$-$40 exhibit C/O ratios slightly exceeding the solar value. Among those with sub-solar C/O ratios, eight objects converge to the model parameter boundary of $\mathrm{C/O} = 0.22$. It should be noted that the uncertainties in C/O could be large due to the linear interpolation errors.

According to the JWST user documentation\footnote{\url{https://jwst-docs.stsci.edu/jwst-calibration-status/miri-calibration-status/miri-lrs-calibration-status}}, the average wavelength uncertainty of the MIRI LRS in slit mode is about 20\,nm, with slightly larger uncertainties near 5\,$\mathrm{\mu m}$. Our fitting results yield average $A_\lambda$ and $C_\lambda$ values from the Sonora Elf Owl models are $A_\lambda=(-4.52\pm1.26)\times10^{-3}$ and $C_\lambda=(44.28\pm12.25)\times10^{-3}$\,$\mathrm{\mu m}$. For the ATMO2020++ models, the average values are $A_\lambda=(-5.68\pm2.19)\times10^{-3}$ and $C_\lambda=(52.15\pm22.16)\times10^{-3}$\,$\mathrm{\mu m}$.
The average wavelength correction parameters derived by the two models are consistent within the 1$\sigma$ uncertainty. For a particular source, the differences in wavelength correction parameters obtained using the two models are primarily due to variations in their model spectra. The two models utilize different opacity databases, which leads to discrepancies in the spectral features.

Moreover, even within the same model, wavelength correction parameters may vary across different sources. This variability could be due to several factors, including errors in data reduction. Additionally, wavelength correction parameters might be influenced by other fitting parameters. The fitting process often involves trade-offs between different parameters, especially when the fit is not perfect across all wavelengths. For example, in the case of ULAS J1029+09, the spectrum between 5.19 and 6\,$\mathrm{\mu m}$ shows that the best-fitting model spectrum derived from the Sonora Elf Owl is systematically lower than the observed spectrum in this wavelength range. To achieve a better match between the model and the observed spectrum, the $\log(R^2/D^2)$ parameter could be increased (shifting the model spectrum upwards), or alternatively, increasing $A_\lambda$ while decreasing $C_\lambda$ (shifting the observed spectrum bluewards) could also work. This is also the reason why we obtained positive values for $A_{\lambda}$ and negative values for $C_{\lambda}$ in some of the sources when using broad prior distributions for the wavelength correction parameters during the preliminary fits (refer to Section~\ref{subsec:method}).

For $\log(R^2/D^2)$, we calculated the corresponding radii relative to $R_{\mathrm{Jup}}$ using the distance $D$ converting from absolute parallax in Table \ref{tab:samples} (see details in Section \ref{subsubsec:radius}). The difference in $\log(R^2/D^2)$ values obtained from the two models is minor, with the Sonora Elf Owl models yielding $\log(R^2/D^2)$ approximately 0.06 larger than those of the ATMO2020++ models.

\begin{figure*}
\centering 
\includegraphics[width=1.\textwidth]{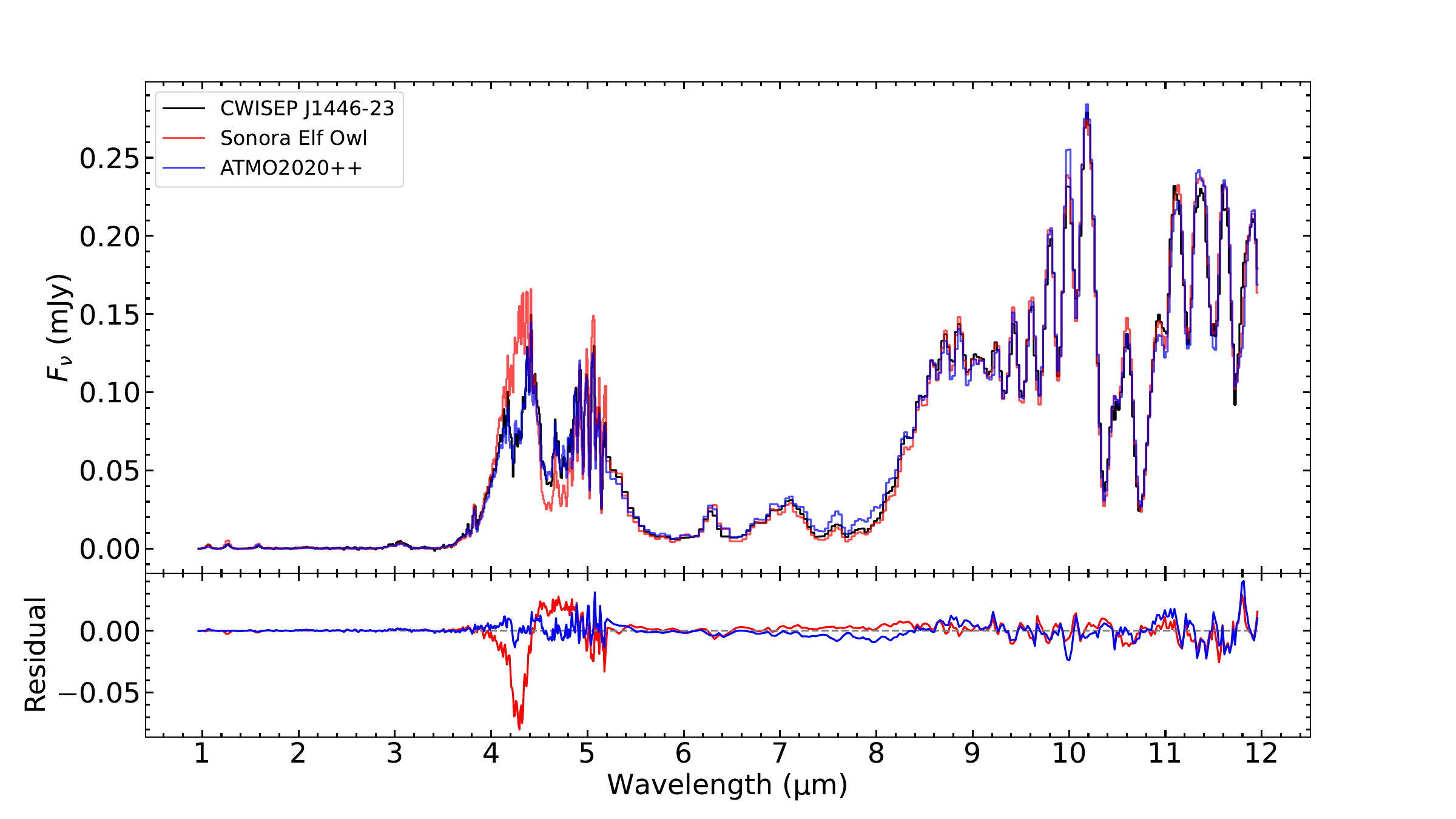}
\includegraphics[width=1.\textwidth]{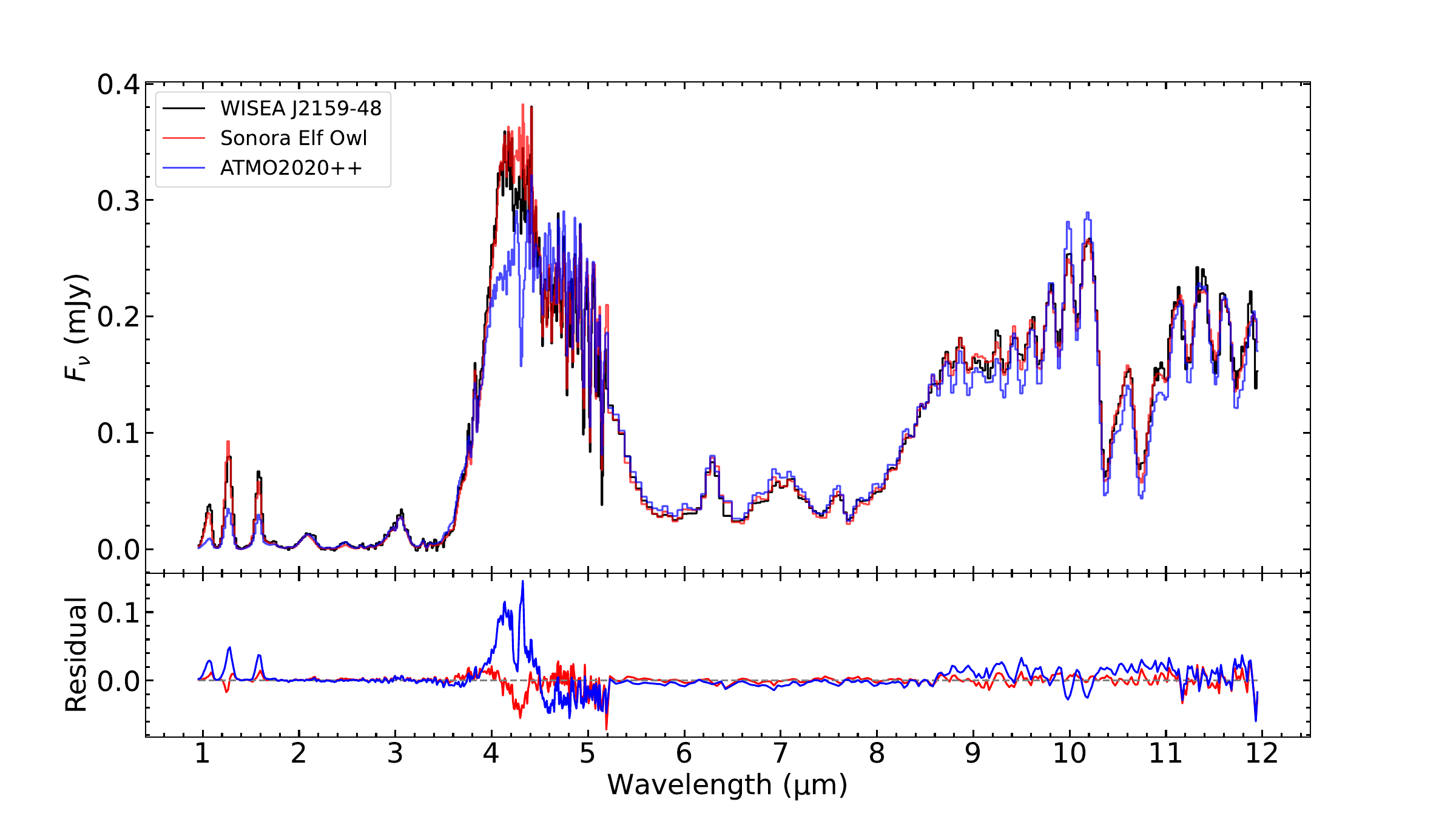}
\caption{Comparisons of the best-fit model spectra derived from Sonora Elf Owl (red) and ATMO2020++ (blue) with the observed spectra (black) of CWISEP J1446$-$23 (top) and WISEA J2159$-$48 (bottom). Both models fit the overall spectra better. Specifically, the results of these two models show that the fitting in the wavelength range of 4--4.5\,$\mathrm{\mu m}$ differs for different objects.\label{fig:obsspec}} 
\end{figure*}

\begin{figure}[t]
\centering 
\includegraphics[width=0.45\textwidth]{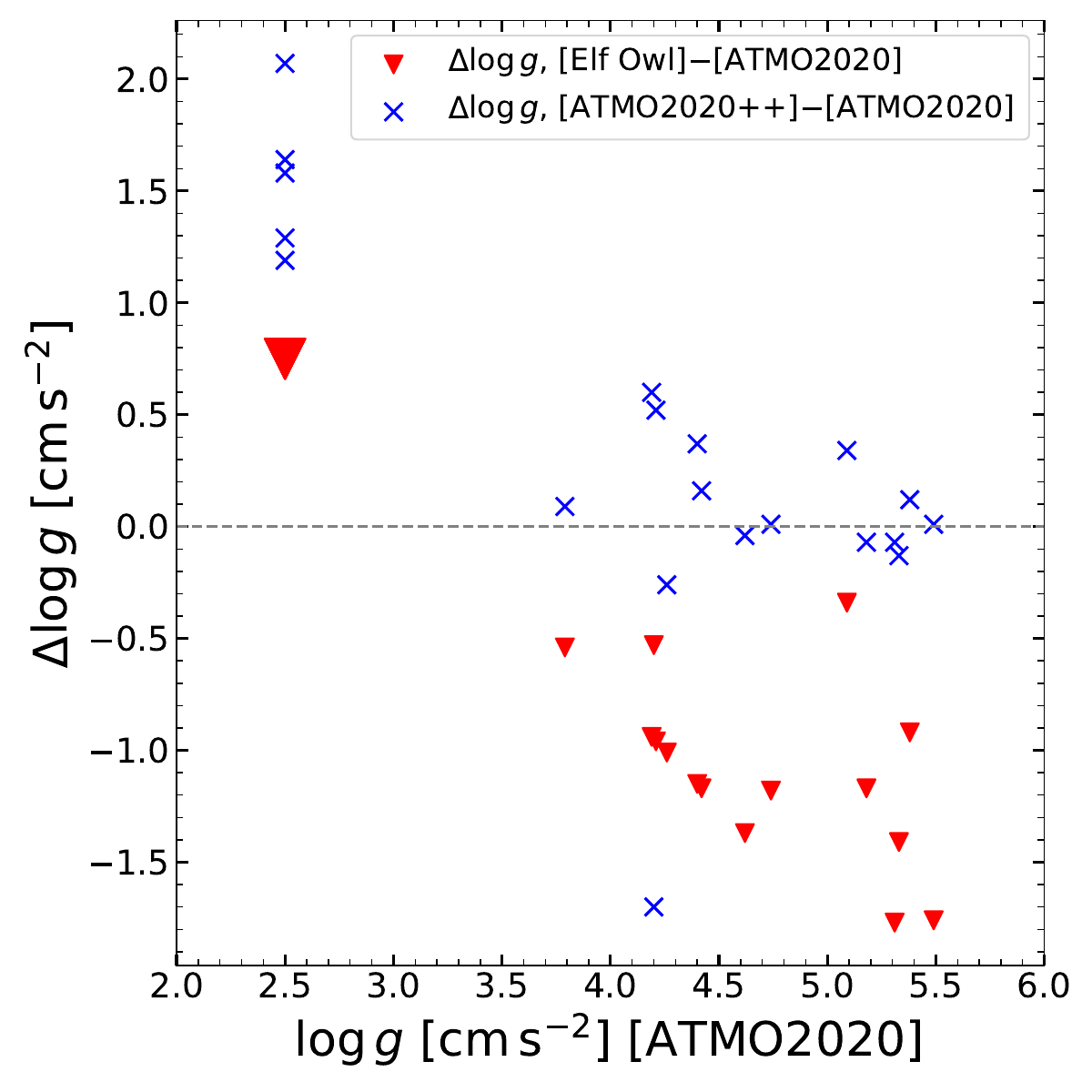}
\caption{Comparisons of $\log g$ derived from the Sonora Elf Owl, ATMO2020++, and ATMO2020 models. The $\log g$ values on the x-axis are from the ATMO2020 models, while the y-axis shows the differences in $\log g$ between the Sonora Elf Owl and ATMO2020 models (red inverted triangles), as well as between the ATMO2020++ and ATMO2020 models (blue crosses). Notably, the largest red inverted triangle at [2.5, 0.75] represents the overlap of five data points.} 
\label{fig:logg_atmo}
\end{figure}

Both models generally fit the overall observed spectra well for all the objects. However, discrepancies remain, mainly in the H$_2$O, CH$_4$, CO$_2$, and CO bands. The differences are especially noticeable in the CO$_2$ band between 4 and 4.5\,$\mathrm{\mu m}$. The ATMO2020++ models always display an absorption feature of PH$_3$ at 4.3\,$\mathrm{\mu m}$, which is less pronounced or absent in the Sonora Elf Owl models. Examples of this can be seen in Figure~\ref{fig:obsspec}, where we present the observed spectra of CWISEP J1446$-$23 and WISEA J2159$-$48 alongside the model spectra from the best-fit parameters of both models. The marginalized posterior distributions for the two objects, using the Sonora Elf Owl and ATMO2020++ models, are shown in Appendix~\ref{app:corner}. For CWISEP J1446$-$23, the ATMO2020++ model spectra fit more appropriately in this region, while for WISEA J2159$-$48, the Sonora Elf Owl model spectra fit better.

The fitting of the spectral region of 4--4.5\,$\mathrm{\mu m}$ is challenging because it possibly includes absorption from both CO$_2$ and PH$_3$. This issue has been reported by \citet{Beiler2024barXiv} in their analysis of four late T and Y dwarfs using various model grids, including the Sonora Elf Owl and ATMO2020++ models. They noted that current substellar atmospheric models tend to predict an overabundance of PH$_3$ and an underabundance of CO$_2$. Specifically, they observed that better spectral agreement could be achieved in this wavelength range if the CO$_2$ abundance in the Sonora Elf Owl models were increased by at least a factor of 100, while the PH$_3$ abundance was reduced by a factor of three. To explain the significant CO$_2$ abundance variation, they proposed that the quench approximation may not accurately predict CO$_2$ levels, suggesting that CO$_2$ remains in chemical equilibrium with CO even after CO quenching occurs, leading to discrepancies in the expected abundance. Additionally, \citet{Leggett2023ApJ} provided another example of possible PH$_3$ overestimation, where spectral fitting for WISE J0359$-$54 using the ATMO2020++ models without PH$_3$ produced a better fit in the 4--4.5\,$\mathrm{\mu m}$ region compared to the ATMO2020++ models with PH$_3$ and the Sonora Bobcat models.

\subsection{Comparison with ATMO2020 models}

From the above analysis, it is evident that the surface gravities obtained by the two models are significantly inconsistent. This discrepancy may arise from differences in the assumptions of the two models, particularly with regard to the convection process. To verify the effect of adiabatic convective processes on surface gravity, we utilize a model with assumptions more similar to those of the Sonora Elf Owl models, which is also the predecessor of the ATMO2020++ models, namely, ATMO2020 \citep{Phillips2020A&A}, to fit our sample. Both the ATMO2020 and Sonora Elf Owl models do not include the reductions in the temperature gradient due to thermochemical instabilities, i.e., the modified effective adiabatic index. But the ATMO2020++ does reduce the temperature gradient of the models by adopting an effective adiabatic index of 1.25, lower than the thermochemical equilibrium value of 1.3--1.4 \citep{Tremblin2015ApJ,Tremblin2016ApJ}.

The ATMO2020 models include both equilibrium and disequilibrium chemistry processes. For consistency, we focus on the components of the ATMO2020 models that utilize disequilibrium chemistry, specifically examining scenarios with ``strong'' and ``weak'' vertical mixing \citep[Figure 1]{Phillips2020A&A}. To identify the best-fit parameters of the model grids, we use the same methods detailed in Section~\ref{subsec:method}. The model spectra of ATMO2020 have not been resampled and retained the original model spectra due to their relatively low resolution.

We subsequently compare the best-fit parameters for each object with those derived using the Sonora Elf Owl and ATMO2020++ models, with a particular emphasis on surface gravity. As illustrated in Figure~\ref{fig:logg_atmo}, the $\log g$ values obtained from the ATMO2020 models are generally larger than those derived from the Sonora Elf Owl models but smaller than those from the ATMO2020++ models. In particular, the $\log g$ values from the ATMO2020 models are closer to those from the ATMO2020++ models than those from the Sonora Elf Owl models. This is expected, given that the ATMO2020++ models share more similarities with the ATMO2020 models.  

Besides, the ATMO2020 models also exhibits five objects with $\log g$ values converging to the lower boundary of $\log g = 2.5$\,[$\mathrm{cm\,s^{-2}}$]. \citet[][Figure 1]{Tremblin2015ApJ} showed that model spectra with reduced temperature gradient exhibit reddening in the $J-H$ color index, and display lower peaks in the $J$ band, on which the $\log g$ measurements are also partially dependent (See Section~\ref{subsec:Partial_spec} for details). These findings suggest that reductions in the temperature gradient due to thermochemical instabilities could influence the determination of surface gravity.

Our analysis using multiple models indicates that while the effective temperature results are relatively consistent, there are notable discrepancies in surface gravity, given that different convective processes are included. Therefore, it is not appropriate to compare results from models with distinct assumptions, as too many variables can interfere with the interpretation of model differences.

\begin{figure*}[t]
\centering 
\includegraphics[width=0.85\textwidth]{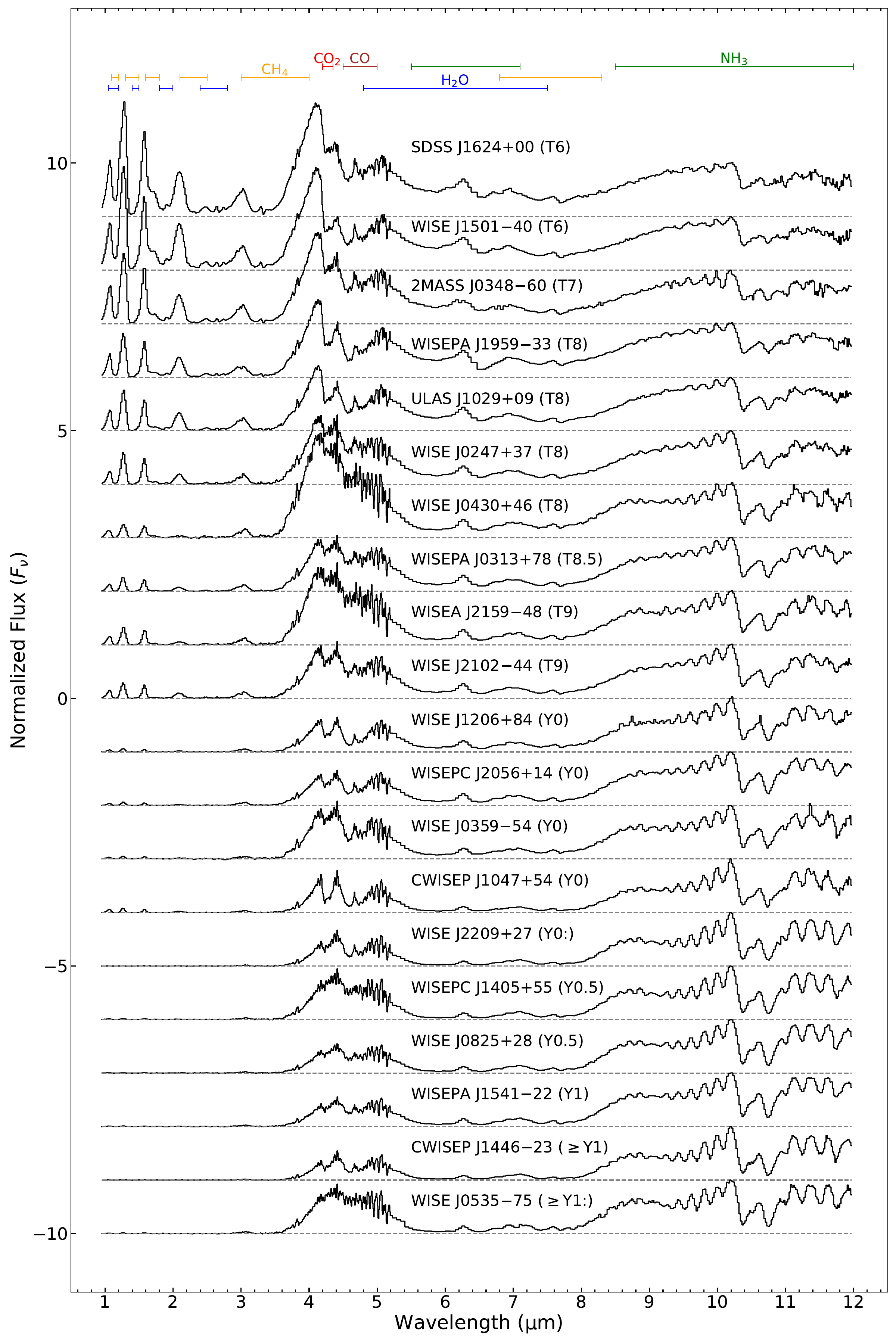}
\caption{The spectra of various objects with spectral types spanning from T6 to $\geq$Y1. The wavelength range is from 0.95 to 12\,$\mathrm{\mu m}$, with main bands of molecular feature listed at the top (same colors indicate the same molecules). The spectral fluxes are normalized at the peak of 10.2\,$\mathrm{\mu m}$.} 
\label{fig:spec_seq}
\end{figure*}

\begin{figure*}[t]
\centering 
\includegraphics[width=1\textwidth]{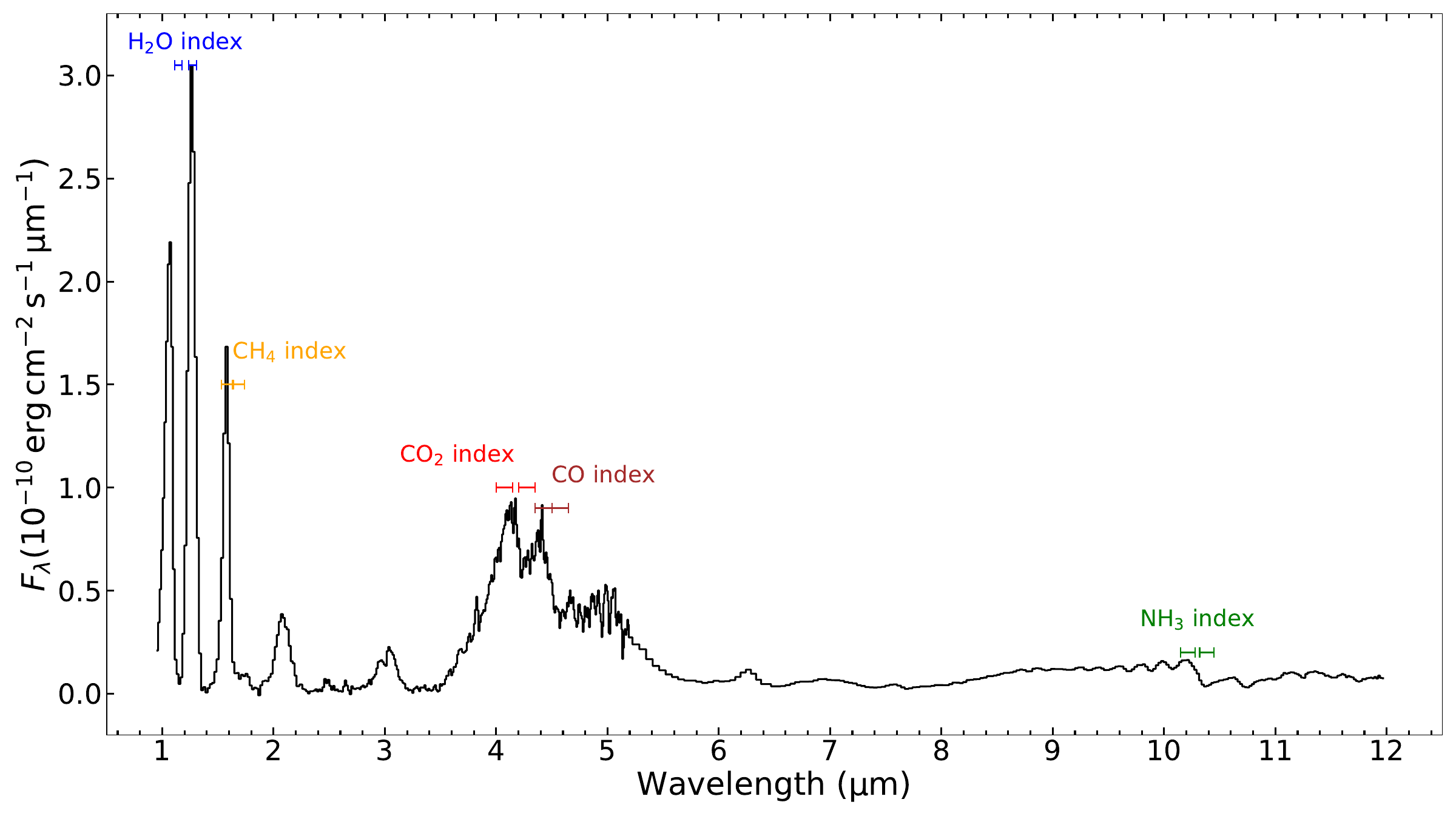}
\caption{An example showing the calculation of spectral indices using specific wavelength intervals from the spectrum of WISE J2102$-$44.} 
\label{fig:spec_index}
\end{figure*}

\subsection{Spectral Sequence and Absorption Features}

Given our extensive sample set, which spans spectral types from T6 to Y1 and beyond, we are capable of investigating how absorption features in the spectra of T and Y dwarfs vary with spectral type. This analysis enables us to discern trends and changes in spectral characteristics as we move across the spectral sequence, providing deeper insights into the atmospheric properties and evolutionary stages of these substellar objects. Figure~\ref{fig:spec_seq} presents the combined spectra for a complete spectral sequence in our sample. The spectral fluxes are normalized at the peak of 10.2\,$\mathrm{\mu m}$.

To compare the relative strengths of the molecular absorption through the spectral type, we define the index of flux ratio for each molecule, as
\begin{equation}
    \text{index}=\frac{\int_{\lambda_1}^{\lambda_2} f_\lambda d\lambda}{\int_{\lambda_3}^{\lambda_4} f_\lambda d\lambda},
\end{equation}
where the wavelength intervals $\lambda_1$--$\lambda_2$ and $\lambda_3$--$\lambda_4$ in the numerator and denominator are of the same length. Figure~\ref{fig:spec_index} presents an example of the wavelength intervals used to calculate the spectral index for each molecule. Specific wavelength intervals and detailed indices through the spectral type for each molecule are listed in Table~\ref{tab:spec_indices}. Due to the very faint absorption features of H$_2$O and CH$_4$ in Y dwarfs, and large flux error within the wavelength region of these molecules in our spectra, the indices of H$_2$O and CH$_4$ for Y dwarfs are not presented.

In the $Y$, $J$, $H$, and $K$ bands, approximately 1--2.3\,$\mathrm{\mu m}$, we observe distinct spectral features shaped by H$_2$O and CH$_4$. These peaks become less pronounced and even invisible in later spectral types, such as Y dwarfs. From Table~\ref{tab:spec_indices}, the relative strengths of H$_2$O and CH$_4$ molecules in these wavelength regions are progressively increasing as the spectral type becomes later. Specifically, the indices of H$_2$O and CH$_4$ for T6 to T8 dwarfs are generally consistent with those calculated by \citet[Table 5 therein]{Geballe2002ApJ}. 

In the wavelength range of 4--5\,$\mathrm{\mu m}$, absorption is primarily dominated by CO, CO$_2$. Notably, the CO$_2$ absorption band around 4.2--4.35\,$\mathrm{\mu m}$ shows significant variation even among objects of the same spectral type, such as T8, T9, and Y0 types in our sample as shown in Table~\ref{tab:spec_indices}. For example, while WISEA J1959$-$33 (T8) and WISEA J0247+37 (T8) both exhibit the CO$_2$ band, the depth and profile of this feature differ significantly. This is also evident from the variation of CO$_2$ indices within the spectral type, where the difference in indices for the same spectral type can be large. We find that indices greater than 1.4 are more clearly outlined in the spectra. In our sample, spectra with deep CO$_2$ absorption features ($\mathrm{index}>1.4$) tend to be found in late T dwarfs, and the only objects with deep features in Y dwarfs are WISE J1206+84 and CWISEP J1047+54. This trend leads the spectral indices of CO$_2$ to become smaller with later spectral types in our sample.

\begin{deluxetable}{lcccccc}

\tablecaption{Spectral Indices and Classification\label{tab:spec_indices}}

\tablehead{\colhead{Object ID} & \colhead{Sp. Type} & \colhead{H$_2$O$^a$} & \colhead{CH$_4$$^b$} & \colhead{CO$_2$$^c$} & \colhead{CO$^d$}& \colhead{NH$_3$$^e$} }

\startdata
SDSS J1624+00&T6&4.58&2.88&1.75&1.67&1.81\\
WISE J1501$-$40&T6&5.55&3.34&2.44&1.71&1.92 \\
2MASS J0348$-$60&T7&6.86&4.60&1.66&1.70&1.95\\
WISEPA J1959$-$33&T8&11.66&5.58&2.14&2.29&2.02\\
ULAS J1029+09&T8&11.30&6.58&1.96&1.99&2.22\\
WISE J0247+37&T8&15.43&8.29&1.22&1.63&2.71\\
WISE J0430+46&T8&16.84&4.88&1.16&1.51&2.87\\
WISEPA J0313+78&T8.5&21.31&10.14&1.14&1.58&2.90\\
WISEA J2159$-$48&T9&19.06&8.42&1.10&1.47&3.09\\
WISE J2102$-$44&T9&25.28&11.78&1.22&1.70&3.34\\
WISE J1206+84&Y0&*&*&1.49&3.05&3.21\\
WISEPC J2056+14&Y0&*&*&1.18&2.12&3.08\\
WISE J0359$-$54&Y0&*&*&1.03&2.06&3.29\\
CWISEP J1047+54&Y0&*&*&1.79&3.41&4.46\\
WISE J2209+27&Y0:&*&*&0.85&2.35&4.81\\
WISEPC J1405+55&Y0.5&*&*&0.78&1.48&4.59\\
WISE J0825+28&Y0.5&*&*&0.83&1.70&3.82\\
WISEPA J1541$-$22&Y1&*&*&0.88&1.69&3.51\\
CWISEP J1446$-$23&$\geq$Y1&*&*&0.96&2.25&4.61\\
WISE J0535$-$75&$\geq$Y1:&*&*&0.80&1.30&3.89\\
\enddata

\tablecomments{The used wavelength intervals for each molecule are:
\tablenotetext{a}{H$_2$O (1.24--1.31)/(1.11--1.18)\,$\mathrm{\mu m}$.}
\tablenotetext{b}{CH$_4$ (1.53--1.63)/(1.64--1.74)\,$\mathrm{\mu m}$.}
\tablenotetext{c}{CO$_2$ (4.00--4.15)/(4.20--4.35)\,$\mathrm{\mu m}$.}
\tablenotetext{d}{CO (4.35--4.50)/(4.50--4.65)\,$\mathrm{\mu m}$.}
\tablenotetext{e}{NH$_3$ (10.15--10.28)/(10.32--10.45)\,$\mathrm{\mu m}$.}
}

\end{deluxetable}

The CO spectral indices show no clear trend of variation with spectral type, remaining mostly around 2 across the range from T6 to Y1. Notably, higher indices ($\mathrm{index} > 3$) are observed only in two sources of the Y0 spectral type. 

The values of spectral indices are influenced by both the molecular absorption cross sections and the gaseous abundances of the respective molecules. Therefore, the observed variations in the spectral indices of CO and CO$_2$ can be attributed to differences in the abundances of these molecules. Changes in the abundances of CO and CO$_2$ may result from vertical mixing processes that quench their abundances in the deeper atmospheric layers. Additionally, the abundances of CO and CO$_2$ are significantly affected by the atmospheric metallicity \citep{Mukherjee2024ApJ}. Therefore, variations in vertical mixing strengths and metallicities could lead to differences in the spectral indices of CO and CO$_2$ even among objects of the same spectral type.

At longer wavelengths around 9--12\,$\mathrm{\mu m}$, the absorption features of NH$_3$ become increasingly apparent and noticeable as one moves towards later spectral types. The indices of NH$_3$ are the same, with larger indices for later spectral types. This trend is also consistent with the results of \citet{Cushing2005ApJ} and \citet{Genaro2022MN}. As noted, the NH$_3$ absorption band at 10.5\,$\mathrm{\mu m}$ first appears at roughly the spectral type L/T transition, and is 
consistently present in $\geq$T2.5 dwarfs.

\subsection{Bolometric Luminosity}

The bolometric luminosity ($L_\mathrm{bol}$) is a crucial parameter that represents the total energy output of a star or astronomical object across all wavelengths, i.e., $L_\mathrm{bol}=4\pi d^2\int_{0}^{\infty} f_\lambda d\lambda$, where the integral term is often referred to as bolometric flux ($F_\mathrm{bol}$). This parameter is essential for determining fundamental stellar properties such as effective temperature, mass, radius, and age. The broadband spectra obtained with JWST allow us to calculate a more accurate bolometric flux. Coupled with the absolute parallaxes listed in Table \ref{tab:samples}, the bolometric luminosity can be derived.

The observed spectra cover the wavelength range from 0.95 to 12\,$\mathrm{\mu m}$. To construct complete spectra, we linearly interpolated the data from zero flux at zero wavelength to the first spectral flux point at 0.95\,$\mathrm{\mu m}$. At the red end, we extended the spectra to $\lambda=\infty$ using the Rayleigh-Jeans tail, given by
\begin{equation} \label{Rayleigh-Jeans tail}
f_\lambda = \frac{2ck_\mathrm{B}T}{\lambda^4}=\frac{C_\mathrm{RJ}}{\lambda^4}.
\end{equation}
The constant $C_\mathrm{RJ}$ is determined by fitting the observed spectra within the wavelength range 11--12\,$\mathrm{\mu m}$.

Using this approach, we calculated the integral values for three regions (i.e., linearly interpolated region, observed spectra, and the Rayleigh-Jeans tail region) to obtain $F_\mathrm{bol}$ and subsequently $L_\mathrm{bol}$. The uncertainties in $L_\mathrm{bol}$ arise from the uncertainties of absolute parallaxes (as listed in Table \ref{tab:samples}) and the flux uncertainties of the observed spectra. The derived $L_\mathrm{bol}$ is presented as $\log(L/\mathcal{L}^N_\odot)$ in Table \ref{tab:mass_age}, where $\mathcal{L}^N_\odot=3.828\times 10^{26}\,\mathrm{W}$ is the intrinsic luminosity of the Sun \citep{Mamajek2015arXiv}.

For comparison, \citet{Beiler2024arXiv} performed precise bolometric luminosity measurements on 23 brown dwarfs using NIRSpec CLEAR/PRISM and MIRI LRS, as well as photometry at 15, 18, and 21\,$\mathrm{\mu m}$. In contrast to our approach, \citet{Beiler2024arXiv} utilized 21\,$\mathrm{\mu m}$ photometry to determine $C_\mathrm{RJ}$ in Equation~(\ref{Rayleigh-Jeans tail}). Comparative analysis reveals that our bolometric luminosity results are generally within 0.05 dex with their calculations. The largest discrepancies of about 0.09 dex come from the object SDSS J1624+00, which they derived $\log(L/\mathcal{L}^N_\odot) = 5.228\pm0.013$.

\begin{figure}[t]
\centering 
\includegraphics[width=0.45\textwidth]{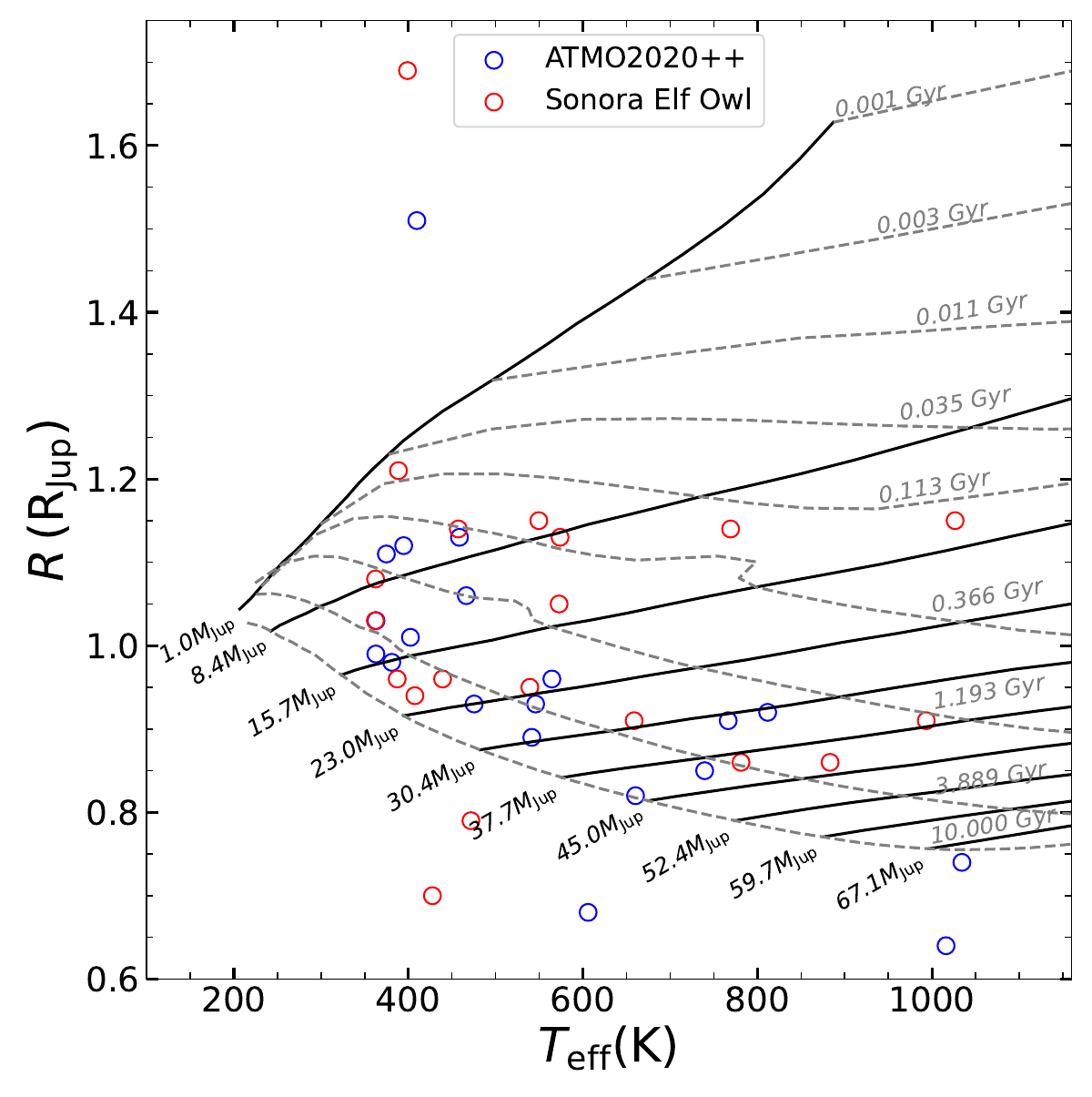}
\caption{Best-fitting parameters of Sonora Elf Owl and ATMO2020++ models from Table \ref{tab:params_elf} and \ref{tab:params_atmo} compared with the evolutionary tracks of \citet{Chabrier2023AA}.} 
\label{fig:Mass_age}
\end{figure}

\begin{deluxetable*}{ccccccccc}

\tablecaption{Model-derived Physical Properties\label{tab:mass_age}}
  \setlength{\tabcolsep}{0.35cm}

\tablehead{\colhead{Object ID} & \colhead{$\log(L/\mathcal{L}^N_\odot)$} & \colhead{$\mathrm{Radius_{evo}}$$^a$} & \colhead{$\mathrm{Radius_{owl}}$}& \colhead{$\mathrm{Mass_{owl}}$} & \colhead{$\mathrm{Age_{owl}}$} & \colhead{$\mathrm{Radius_{atmo}}$} & \colhead{$\mathrm{Mass_{atmo}}$} & \colhead{$\mathrm{Age_{atmo}}$} \\ 
\colhead{} & \colhead{} & \colhead{($R_\mathrm{Jup}$)}& \colhead{($R_\mathrm{Jup}$)} & \colhead{($M_\mathrm{Jup}$)} & \colhead{(Gyr)} & \colhead{($R_\mathrm{Jup}$)} & \colhead{($M_\mathrm{Jup}$)} & \colhead{(Gyr)} }

\startdata
WISE J0247+37&$-5.83\pm0.03$&0.76--1.19&$0.91\pm0.03$&29.3 & 3.07&$0.82\pm0.03$&43.6 & 9.65\\
WISEPA J0313+78&$-6.00\pm0.02$&0.91--1.20&$1.03\pm0.02$&11.4 & 3.03&$0.96\pm0.02$&21.5 & 2.43\\
2MASS J0348$-$60&$-5.29\pm0.01$&0.93--1.18&$0.91\pm0.01$&36.6 & 1.33&$0.92\pm0.01$&31.0 & 1.77\\
WISE J0359$-$54&$-6.39\pm0.03$&0.76--1.18&$1.05\pm0.03$&13.7 & 0.99&$0.93\pm0.03$&23.1 & 5.17\\
WISE J0430+46&$-6.21\pm0.03$&0.79--1.21&$0.86\pm0.03$&42.2 & 2.88&$0.68\pm0.02$&*  &*\\
WISE J0535$-$75&$-6.26\pm0.03$&0.84--1.22&$1.69\pm0.05$&*  &*  &$1.51\pm0.04$&*  &*\\
WISE J0825+28&$-6.68\pm0.03$&0.81--1.21&$1.14\pm0.01$&6.4  &0.38 &$1.11\pm0.01$&6.6  &0.88\\
ULAS J1029+09&$-5.61\pm0.02$&0.87--1.21&$0.86\pm0.02$&39.7 & 3.73&$0.85\pm0.02$&40.4 & 4.75\\
CWISEP J1047+54&$-6.60\pm0.07$&0.85--1.21&$0.94\pm0.07$&20.6 & 6.73&$0.98\pm0.07$&15.9 & 5.01\\
WISE J1206+84&$-6.30\pm0.03$&0.86--1.21&$1.21\pm0.03$&1.9  &0.08&$1.06\pm0.03$&11.2 & 1.19\\
WISEPC J1405+55&$-6.64\pm0.02$&0.91--1.19&$1.15\pm0.02$&13.6 & 0.16&$1.01\pm0.02$&13.8 & 3.18\\
CWISEP J1446$-$23&$-6.85\pm0.05$&0.86--1.21&$1.08\pm0.05$&8.0  &1.47&$0.99\pm0.05$&14.6 & 5.10\\
WISE J1501$-$40&$-5.30\pm0.03$&0.76--1.19&$0.70\pm0.02$&*  &*&$0.64\pm0.02$&*  &*\\
WISEPA J1541$-$22&$-6.58\pm0.03$&0.83--1.22&$1.14\pm0.01$&10.8 & 0.18&$1.12\pm0.01$&6.4  &0.67\\
SDSS J1624+00&$-5.14\pm0.01$&0.93--1.17&$0.79\pm0.01$&*  &*&$0.74\pm0.01$&*  &*\\
WISEPA J1959$-$33&$-5.50\pm0.02$&0.83--1.22&$0.96\pm0.02$&18.0 & 6.11&$0.91\pm0.02$&31.5 & 2.21\\
WISEPC J2056+14&$-6.29\pm0.02$&0.90--1.20&$1.15\pm0.02$&7.3  &0.26&$1.13\pm0.02$&6.9  &0.45\\
WISE J2102$-$44&$-6.04\pm0.02$&0.77--1.20&$0.95\pm0.02$&22.1 & 3.02&$0.93\pm0.02$&24.5 & 3.70\\
WISEA J2159$-$48&$-6.09\pm0.03$&0.90--1.20&$0.96\pm0.03$&19.1 & 4.41&$0.89\pm0.03$&29.6 & 6.02\\
WISE J2209+27&$-6.82\pm0.03$&0.86--1.21&$1.13\pm0.01$&8.7  &0.31&$1.03\pm0.01$&11.4 & 3.01\\
\enddata

\tablecomments{The physical properties (Radius, mass, and age), denoted with subscripts ``owl'' and ``atmo'', are calculated using the fitted parameters from Sonora Elf Owl and ATMO2020++ models, respectively. The radius is computed using the fitted parameter $\log(R^2/D^2)$ and the absolute parallaxes in Table \ref{tab:samples}. Masse and age are determined from the effective temperature and radius. Specifically, Asterisks (*) indicate values are beyond the boundaries of the evolutionary tracks.
\tablenotetext{a}{This radius is obtained using the linearly-interpolated evolutionary tracks of \citet{Chabrier2023AA}, based on the bolometric luminosity listed in column 2, and assuming an age range of 0.1 to 10\,Gyr.}}

\end{deluxetable*}

\subsection{Properties From Evolutionary Models} \label{subsec:massage}

\subsubsection{Evolutionary Models} \label{subsubsec:evo_models}

The parameters derived from atmospheric models are crucial in determining the masses and ages of brown dwarfs through evolutionary models. Since the atmospheric model-fitting results discussed in Section~\ref{sec:modelfit} do not impose physical constraints on those parameters, comparing these values with evolutionary models can validate their physical plausibility. 

To this end, we use the evolutionary models of \citet{Chabrier2023AA}. These evolutionary models utilize the ATMO2020 atmospheric models and incorporate the equation of state for dense hydrogen-helium mixtures, as well as interactions between hydrogen and helium species during the evolution of very low-mass dwarfs. Two sets of evolutionary tracks are calculated, representing ``strong'' and ``weak'' vertical mixing scenarios (see \citealt{Phillips2020A&A}, Figure 1). The results from these two evolutionary tracks are highly consistent. Thus, we arbitrarily adopted the evolutionary tracks with ``strong'' vertical mixing strength.

\subsubsection{Radius} \label{subsubsec:radius}

We determined the radii of our sample using two methods. The first method involves directly calculating the radius using the fitted parameters $\log(R^2/D^2)$ from the two models and the absolute parallaxes in Table \ref{tab:samples}. The second method is to use linearly interpolated evolutionary tracks and the calculated bolometric luminosity to obtain the range of radii under the evolutionary tracks, assuming ages between 0.1 and 10\,Gyr.

The results of the radii from both methods are presented in Table~\ref{tab:mass_age}. The radii derived with the fitted parameters from the Sonora Elf Owl models are on average 0.07\,$R_{\mathrm{Jup}}$ larger than those from the ATMO2020++ models, related to the $\log(R^2/D^2)$ parameters mentioned in Section~\ref{subsec:model_fit_results}. Most radii obtained through fitted parameters range from approximately 0.8 to 1.2\,$R_{\mathrm{Jup}}$, consistent with those obtained through the evolutionary tracks. However, there are three objects for which the radii obtained by these two methods do not coincide---WISE J0535$-$75, WISE J1501$-$40, and SDSS J1624+00.

A particular object, WISE J0535$-$75, shows an unusually large radius. The radii obtained from both models, $R_\mathrm{owl}=1.69\pm0.05$\,$R_{\mathrm{Jup}}$ and $R_\mathrm{atmo}=1.51\pm0.04$\,$R_{\mathrm{Jup}}$, are much larger than the radius derived from the evolutionary tracks assuming an age of 0.1\,Gyr, i.e., $R_\mathrm{evo}=1.22$\,$R_{\mathrm{Jup}}$. For comparison, the object WISEPC1405+55, with similar $T_\mathrm{eff}$ and $\log g$ to WISE J0535$-$75, has a radius of about $1\,R_{\mathrm{Jup}}$. 

These discrepancies might be due to the objects being in a different evolutionary phase or having peculiar atmospheric properties that are not well represented by the standard evolutionary models used in this analysis. Furthermore, uncertainties in parallax measurements or potential errors in the atmospheric model assumptions might contribute to this deviation. For instance, if the parallax measurement is slightly underestimated, it would lead to a larger inferred radius.

\subsubsection{Mass and Age} \label{subsubsec:mass_age}

Typically, the mass and age of an object are determined using effective temperature and surface gravity derived from model parameters. In our model-fitting results, the effective temperatures from the two models show good agreement, while there is a notable discrepancy in the surface gravity values. This discrepancy leads to significant variations in the derived masses and ages, complicating the accurate determination of their true properties. Given that the distances of these objects are mostly within 20\,pc, the parallax errors are relatively small. Therefore, with the linearly interpolated evolutionary tracks, we have determined the objects' masses and ages using the effective temperatures from the two models and the radii calculated by the fitting parameters $\log(R^2/D^2)$ with the absolute parallax values. The results are presented in Table~\ref{tab:mass_age} and Figure~\ref{fig:Mass_age}.

The discrepancy in radii derived from the two models directly impacts the resultant masses and ages of the objects in our sample. According to evolutionary tracks, the larger radii obtained from the Sonora Elf Owl models lead to smaller estimated masses and ages compared to those derived from the ATMO2020++ models. Specifically, the average mass difference between the two models is 5.5\,$M_{\text{Jup}}$, and the average age difference is 1.8\,Gyr.

The mass distribution of our sample ranges from 2 to 45\,$M_{\text{Jup}}$, with most objects having masses below 30\,$M_{\text{Jup}}$. In particular, five objects, all of which are Y dwarfs, have masses from both the two models less than the theoretical boundary of 13\,$M_{\text{Jup}}$ \citep{Burrows2001RvMP}. Focusing on the Y dwarfs, the average mass using the Sonora Elf Owl models is 10.1\,$M_{\text{Jup}}$, while the ATMO2020++ models give an average mass of 12.2\,$M_{\text{Jup}}$. These masses range from 2 to approximately 20\,$M_{\text{Jup}}$, consistent with the conclusions of \citet{Leggett2017ApJ}.

The age distribution of our sample spans from 0.1 to 10\,Gyr, with the majority of objects being younger than 6\,Gyr. For Y dwarfs, the average age for the Sonora Elf Owl models is 1.17\,Gyr, and 2.74\,Gyr for the ATMO2020++ models. These ages range from 0.1 to 6.7\,Gyr, slightly less than the findings of \citet{Leggett2017ApJ}.

Compared to deriving mass and age using effective temperature and surface gravity, this approach, which utilizes absolute parallaxes and $\log(R^2/D^2)$, offers more consistent properties from the parameters of the two models and reduces uncertainties related to model-specific assumptions and limitations, especially when there are systematic differences in surface gravity derived from the two models.

\section{Discussion} \label{sec:discuss}

\subsection{Atmospheric Parameters Using Partial Spectra} \label{subsec:Partial_spec}

\begin{figure*}[t]
\centering 
\includegraphics[width=1\textwidth]{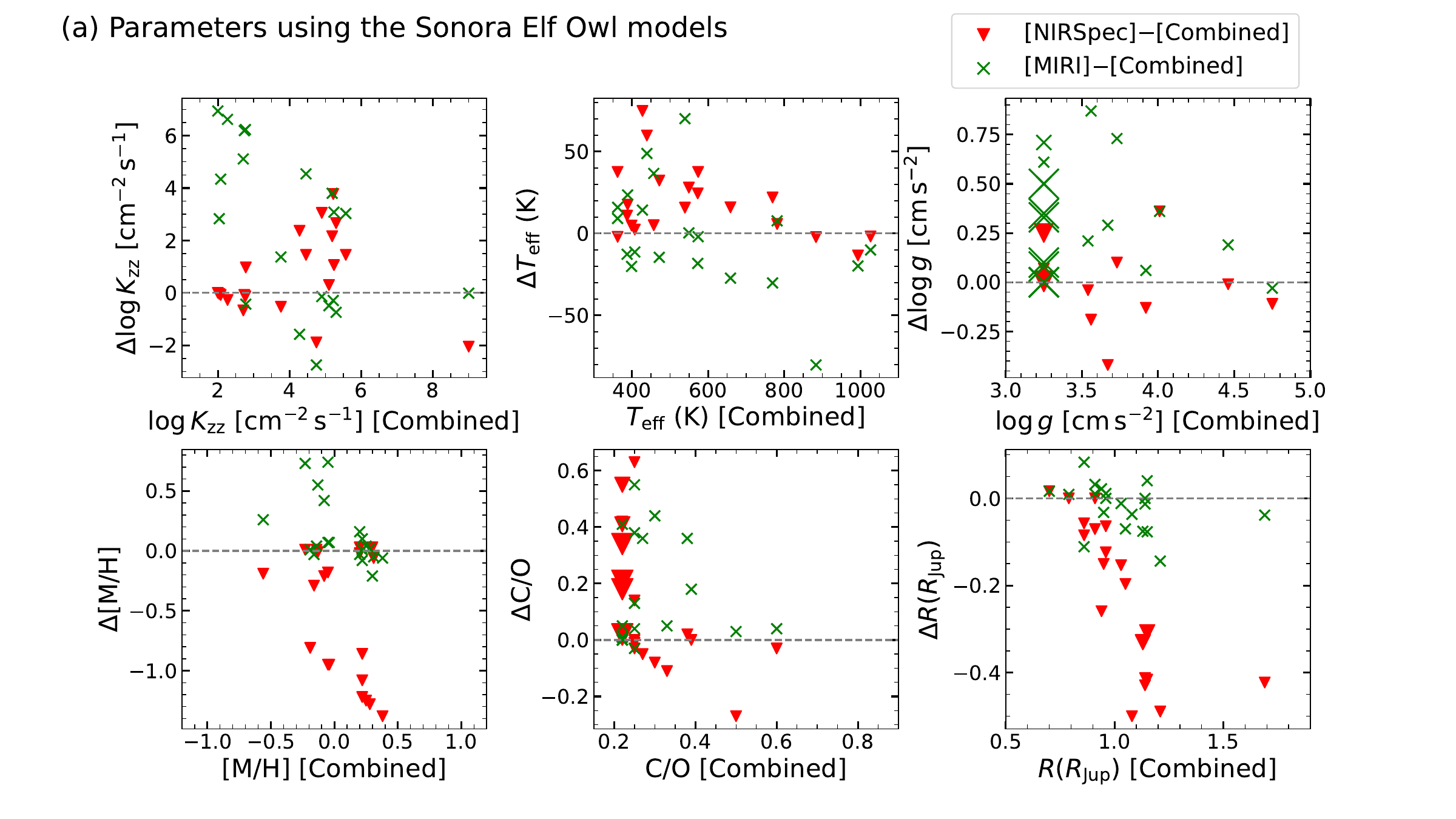}
\includegraphics[width=0.67\textwidth]{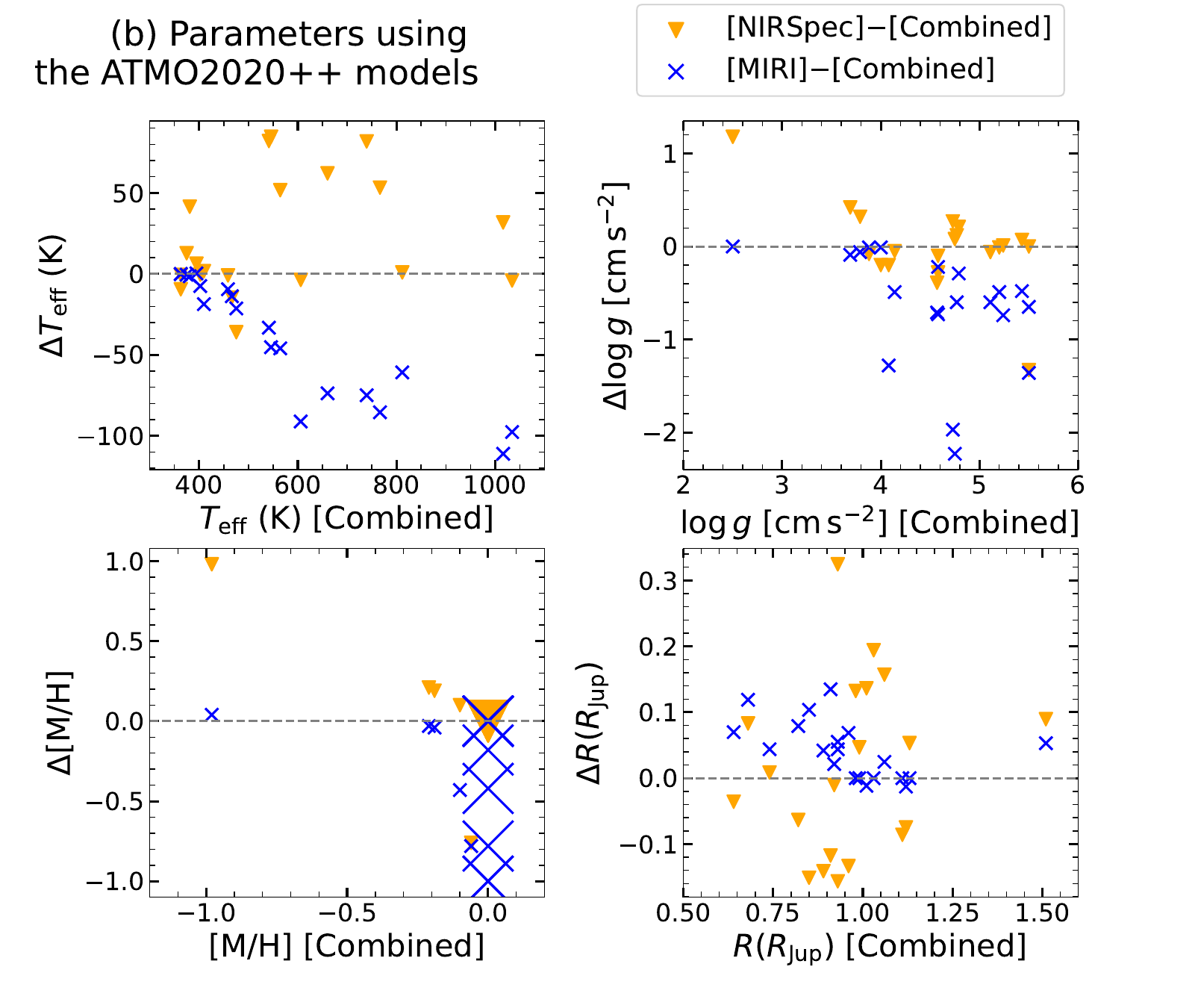}
\caption{Comparisons of parameters derived by the Sonora Elf Owl (Figure a) and ATMO2020++ (Figure b) models with spectra from different instruments versus combined spectra. The parameters on the x-axis are obtained from the combined spectra (NIRSpec+MIRI), i.e., the values listed in Table \ref{tab:params_elf}, \ref{tab:params_atmo}, and \ref{tab:mass_age}. The y-axis shows the differences between parameters derived from NIRSpec-only and combined spectra (inverted triangles), as well as differences between MIRI-only and combined spectra (crosses). Larger data points indicate a greater number of overlapping objects at that position.} 
\label{fig:partial_spec}
\end{figure*}

Accurately determining the atmospheric parameters of brown dwarfs is crucial for understanding their physical properties and evolutionary status. Comprehensive spectral data across a wide range of wavelengths can be used to derive general parameters more effectively. In our previous analysis, we utilized the combined spectra from the NIRSpec CLEAR/PRISM and MIRI LRS instruments to achieve a detailed characterization of atmospheric parameters, allowing for comprehensive model fitting and accurate calculation of bolometric luminosity.

However, in many practical scenarios, only partial spectral data may be available due to observational constraints or limitations to specific instruments. This subsection explores the discrepancies in deriving atmospheric parameters using partial spectra---specifically, data obtained solely from NIRSpec CLEAR/PRISM or MIRI LRS---and compares these results with those obtained from the combined spectral analysis (NIRSpec+MIRI).

By focusing on data obtained separately from NIRSpec CLEAR/PRISM and MIRI LRS, we aim to assess the potential for these instruments to independently provide reliable atmospheric parameter estimates. This analysis can help understand the limitations and capabilities of each instrument when used in isolation, providing insights into how well partial spectra can approximate results obtained from the full spectral range. Furthermore, it informs additional observational strategies, particularly in cases where obtaining full spectral coverage is not feasible. This analysis is crucial for optimizing the use of JWST's instruments and for interpreting archival data where complete spectra are not always accessible.

For the analysis using only the spectra from NIRSpec CLEAR/PRISM, the methodology described in Section~\ref{subsec:method} was employed, with the exception that two wavelength correction parameters were excluded. All other parameter settings remained consistent with those outlined in Section~\ref{subsec:method}. While in the analysis using only the spectra from MIRI LRS, the same methodology as processing the combined spectra was applied.

In the following analysis, we focus on the effect of different bands of spectra on the fitted atmospheric parameters. For the sake of brevity, parameters obtained using only the spectra of NIRSpec CLEAR/PRISM are referred to as NIRSpec-only parameters. Similarly, parameters obtained using only the spectra of MIRI LRS are referred to as MIRI-only parameters, while those derived from the combined spectra are termed combined parameters.

\subsubsection{Comparison with Effective Temperature}

Figure~\ref{fig:partial_spec} illustrate the atmospheric parameters derived from NIRSpec-only and MIRI-only observations using the Sonora Elf Owl and ATMO2020++ models, compared to those obtained from combined spectra. The x-axis values represent the parameters derived from the combined spectra, while the y-axis shows the differences between NIRSpec-only and combined values, as well as between MIRI-only and combined values, distinguished by different colors and symbols.

From the comparisons of effective temperatures in Figure~\ref{fig:partial_spec}, the effective temperatures derived from a single instrument show good agreement with those obtained from combined spectra. Those $T_\mathrm{eff}$ from the ATMO2020++ models reveal that for combined $T_\mathrm{eff}$ above approximately 500 K, the NIRSpec-only $T_\mathrm{eff}$ tend to be slightly higher, while the MIRI-only $T_\mathrm{eff}$ are generally lower. This trend is less pronounced for those from the Sonora Elf Owl models, but similar patterns are observed: NIRSpec-only values are higher in the low $T_\mathrm{eff}$ range, and MIRI-only values are lower in the high $T_\mathrm{eff}$ range.

This discrepancy can be attributed to the fact that the shorter wavelength range covered by NIRSpec is more influenced by the hotter regions of a brown dwarf's atmosphere, being sensitive to higher temperature indicators such as water and methane features. These features dominate in warmer brown dwarfs and can lead to an overestimation of the effective temperature when analyzed in isolation. Conversely, the mid-infrared range covered by MIRI is more sensitive to cooler temperature indicators such as ammonia, resulting in an underestimation of the effective temperature when not complemented by shorter wavelength observations.

The differing effective temperatures derived from NIRSpec-only and MIRI-only spectra emphasize the importance of utilizing a broad wavelength range to accurately characterize the effective temperature of brown dwarfs. Despite these differences, the variations in effective temperature between single-instrument spectra and combined spectra are not substantial. The effective temperatures in our sample obtained using either NIRSpec or MIRI generally remain within a reliable range (difference within $\pm50$\,K), demonstrating the capability of these instruments to provide valuable effective temperature estimates even when used independently.

\subsubsection{Comparison with Surface Gravity}

For those surface gravities derived from the Sonora Elf Owl models, there is a strong correlation between the values obtained using only NIRSpec and those using the combined spectra, i.e., the differences between NIRSpec-only and combined $\log g$ are mostly within 0.25 dex. This correlation, however, is weaker for the values derived from MIRI-only data.

Similarly, for those surface gravities obtained by using the ATMO2020++ models, a strong linear correlation is observed between NIRSpec-only and combined values. Despite two points with significant deviations, the NIRSpec-only measurements at these two points appear more reasonable. For example, one point shows a combined surface gravity value of 2.5 dex, which exceeds the physical boundary in the evolutionary tracks, while the corresponding NIRSpec-only measurement is more physically plausible. However, MIRI-only values are generally smaller than those from combined spectra, with some objects showing a combined surface gravity of 4 to 5 dex but MIRI-only values of less than 3 dex.

Examining the spectra of the Sonora Elf Owl and ATMO2020++ models within the 1--12\,$\mathrm{\mu m}$ wavelength range reveals that surface gravity variations impact spectra in three prominent regions (measured in spectral flux units of Jansky): (1) the H$_2$O and CH$_4$ absorption feature at 1--2.5\,$\mathrm{\mu m}$, where higher surface gravity is associated with fainter fluxes; (2) the CO$_2$ and CO absorption features at 4--5\,$\mathrm{\mu m}$, where higher gravity results in brighter fluxes, with more pronounced variations in CO$_2$ compared to CO; and (3) the NH$_3$ absorption spectra at 8.5--12\,$\mathrm{\mu m}$, with flux decreasing throughout the wavelength range with increasing surface gravity.

Regions (1) and (2) are both covered by the NIRSpec CLEAR/PRISM spectra, which explains why NIRSpec-only values are closer to those derived from the combined spectra, thus being more consistent. In contrast, only region (3) is covered by the MIRI LRS spectra. The spectra in this region vary globally with surface gravity, leading to a degeneracy between surface gravity and effective temperature. This degeneracy complicates the determination of surface gravity when using only MIRI LRS spectra.

In order to investigate the degeneracy between surface gravity and effective temperature in this band, we also performed another fit for only MIRI spectra, using the ATMO2020++ models with a fixed combined $T_\mathrm{eff}$. The results show that the average difference between combined and MIRI-only $\log g$ with fixed combined $T_\mathrm{eff}$ is merely 0.05 dex. However, when $T_\mathrm{eff}$ is allowed to vary as a free parameter, the average difference in $\log g$ increases significantly to 0.45 dex.

\subsubsection{Comparison with Metallicity}

The metallicities derived using the Sonora Elf Owl models, based on combined spectra, mostly fall within the near-solar range of $-$0.5 to 0.5. The metallicities obtained from MIRI-only spectra show strong agreement with those derived from the combined spectra, with 15 out of 20 objects showing a difference of less than $\pm0.2$ dex. However, this consistency is not observed in the NIRSpec-only metallicities, with some of the NIRSpec-only metallicities falling below $-$0.5 and even reaching as low as $-$1.

The metallicity range derived from the ATMO2020++ models differs from that obtained from the Sonora Elf Owl models, spanning values between $-1$ and 0. Most combined metallicities are clustered around zero. This trend is also evident in the NIRSpec-only metallicities, with only one object not showing a metallicity equal to zero. In contrast, most MIRI-only metallicities are smaller than the combined metallicities, with values more evenly distributed between $-1$ and 0.

Variations in metallicity significantly impact spectra in regions with strong molecular absorption. According to \citet[Figure~7]{Mukherjee2024ApJ}, the H$_2$O and CH$_4$ bands (1--2.5\,$\mathrm{\mu m}$), the CO$_2$ band (4.2--4.35\,$\mathrm{\mu m}$), and the CO band (4.5--5\,$\mathrm{\mu m}$) are the primary regions impacted by metallicity, with CO$_2$ and CO bands showing more significant spectral variations.

Accurately determining metallicity is challenging because variations in these spectral regions are also influenced by effective temperature and surface gravity. Furthermore, the observed spectra in the CO$_2$ and CO bands often do not match well with the forward model spectra, leading to significant uncertainties in the determination of metallicity. The metallicity is highly dependent on the observed spectra, explaining why individual instruments and combined spectra yield different metallicities.

\subsubsection{Comparison with $\log K_\mathrm{zz}$, C/O, and radius}

The vertical eddy diffusion parameter $K_\mathrm{zz}$ and C/O ratio are two atmospheric parameters in the Sonora Elf Owl models. As shown in Figure~\ref{fig:partial_spec}, both $\log K_\mathrm{zz}$ and C/O values derived from either NIRSpec-only or MIRI-only spectra exhibit significant deviations compared to those obtained from combined spectra.

According to \citet[Figure~7]{Mukherjee2024ApJ}, spectral variations due to $K_\mathrm{zz}$ mainly focus on the CH$_4$, CO$_2$ and CO bands. The CO band shows more rapid changes with varying $K_\mathrm{zz}$ compared to the CO$_2$ band. This theoretically helps to break the degeneracy between $K_\mathrm{zz}$ and metallicity. However, the observed spectra for the CO$_2$ and CO bands often do not match well with model spectra, leading to substantial uncertainties in parameters associated with these regions.

The varying C/O ratio only presents the difference in CH$_4$ bands. The CO and CO$_2$ bands, however, are nearly the same with different C/O ratios. This makes the C/O ratio more sensitive to the CH$_4$ bands. But the CH$_4$ bands are also one of the poorly fitted regions when using the combined spectra, with 16 out of 20 objects having significant residuals at the CH$_4$ bands. Such an issue may cause differences between the C/O ratio from the spectra of individual instruments and those from the combined spectra.

For the radii from the parameter $\log (R^2/D^2)$, as illustrated in Figure~\ref{fig:partial_spec}, the differences between the MIRI-only and the combined radii, derived from the Sonora Elf Owl models, are generally smaller compared to the differences observed between the NIRSpec and the combined values. Specifically, the deviations in MIRI-only radii from the combined values are mostly within $\pm 0.1$\,$R_\mathrm{Jup}$, whereas the NIRSpec values tend to be systematically lower than the combined radii, with some discrepancies reaching up to $-0.5$\,$R_\mathrm{Jup}$. The results obtained using the ATMO2020++ models are similar to those from the Sonora Elf Owl models, where the MIRI-only radii show better consistency with the combined values, with differences mostly constrained within $\pm 0.1$\,$R_\mathrm{Jup}$. In contrast, the NIRSpec-only radii exhibit greater scatter.

\subsection{Atmospheric Parameters in Previous Works}

In the previous sections, we have analyzed the model-fitting results for our sample in general. Since some objects have been studied for atmospheric parameters, we provide a detailed comparison of our results with those reported in the literature. To facilitate this comparison, parameters derived from the Sonora Elf Owl models are denoted with the subscript ``owl'', and parameters derived from the ATMO2020++ models are denoted with the subscript ``atmo''.

\subsubsection{Effective Temperature}

From T6 to Y1 spectral types, the effective temperatures derived from both models generally agree well with previous studies. For instance, for the T6 dwarf SDSS J1624+00, we find effective temperatures of $T_{\mathrm{eff,owl}} = 1024.46 \pm 0.29\,\text{K}$ and $T_{\mathrm{eff,atmo}} = 1031.31 \pm 0.29\,\text{K}$, which are consistent with the $T_{\mathrm{eff}} = 980 \pm 163\,\text{K}$ reported by \citet{DelBurgo2009A&A}. Our results are slightly higher, but the agreement is generally within the expected uncertainties.

For the T7 dwarf 2MASS J0348$-$60, \citet{Tannock2021AJ} reported an effective temperature of $T_{\mathrm{eff}} = 880 \pm 110\,\text{K}$. Our results yield $T_{\mathrm{eff,owl}} = 882.71 \pm 0.08\,\text{K}$ and $T_{\mathrm{eff,atmo}} = 810.97 \pm 0.01\,\text{K}$. The Sonora Elf Owl models agree well with the previous estimate, whereas the ATMO2020++ models provide a relatively lower temperature.

One of the coldest brown dwarfs in our sample, WISE J0535$-$75, is thought to have an effective temperature in the range of 360--390\,K according to \citet{Leggett2017ApJ}, or 400--450\,K as suggested by \citet{Schneider2015ApJ}. Our models give $T_{\mathrm{eff,owl}} = 398.97 \pm 0.35\,\text{K}$ and $T_{\mathrm{eff,atmo}} = 409.71^{+0.17}_{-0.18}\,\text{K}$, demonstrating that both models provide reasonable fits to the observed data for the very cold brown dwarf.

\subsubsection{Surface Gravity}

The surface gravities derived from the two models show systematic differences. Generally, the surface gravities obtained in the literature align more closely with those from the ATMO2020++ models. For example, \citet{Tannock2021AJ} found $\log g = 5.1 \pm 0.3$\,[$\mathrm{cm\,s^{-2}}$] for 2MASS J0348$-$60, which is in better agreement with our results of $\log g_{\mathrm{atmo}} \approx 5.23$\,[$\mathrm{cm\,s^{-2}}$] compared to the results of $\log g_{\mathrm{owl}} \approx 3.54$\,[$\mathrm{cm\,s^{-2}}$].

This trend is also observed in other objects. For WISEPC J2056+14, \citet{Miles2020AJ} derived $\log g = 4.4 \text{--} 5.0$\,[$\mathrm{cm\,s^{-2}}$] using semi-empirical methods from the Sonora Bobcat evolutionary models, which assumes ages between 1 and 10\,Gyr. Our results show $\log g_{\mathrm{atmo}} \approx 4.12$\,[$\mathrm{cm\,s^{-2}}$] and $\log g_{\mathrm{owl}} \approx 3.25$\,[$\mathrm{cm\,s^{-2}}$], with the former one being closer to the value in the literature.

For WISE J0359$-$54, \citet{Beiler2023ApJ} found $\log g = 3.25$\,[$\mathrm{cm\,s^{-2}}$] using the Sonora Bobcat and Sonora Cholla models. This result aligns with that of Sonora Elf Owl models, while the ATMO2020++ models yield a higher $\log g_{\mathrm{atmo}} \approx 4.56$\,[$\mathrm{cm\,s^{-2}}$], which is more consistent with the semi-empirical measurements provided by \citet{Beiler2023ApJ}, suggesting a range of 4.25 to 5.0\,[$\mathrm{cm\,s^{-2}}$].

\subsubsection{Wavelength Correction Parameters}

In our analysis, the adopted wavelength correction parameters using the Sonora Elf Owl models are $A_\lambda=(-4.52\pm1.26)\times10^{-3}$ and $C_\lambda=(44.28\pm12.25)\times10^{-3}$\,$\mathrm{\mu m}$, which means the relationship between the observed wavelength of MIRI LRS spectra and the corrected wavelength from Equation~(\ref{eq:wave_cor}) is
\begin{equation}
\lambda_\mathrm{cor}=\lambda_\mathrm{MIRI}+\Delta\lambda\approx \lambda_\mathrm{MIRI}-0.00452\lambda_\mathrm{MIRI}+0.0443
\end{equation}

However, in dealing with the spectra of WISE J0359$-$54, \citet{Beiler2023ApJ} measured the wavelength differences at several peaks between the MIRI observed spectrum and a 500\,K spectrum of the Sonora Cholla models. They used a linear function to fit the wavelength deviation of the MIRI observed spectra, resulting in the equation $\Delta\lambda = 0.0106\lambda_\mathrm{MIRI} - 0.120$. The coefficients 0.0106 and $-$0.120 correspond to our wavelength correction parameters $A_\lambda$ and $C_\lambda$, respectively.

This comparison reveals that our derived wavelength correction relationship differs from the results of \citet{Beiler2023ApJ}. Specifically, their relationship indicates a wavelength deviation of approximately 70\,nm at 5\,$\mathrm{\mu m}$, whereas our results demonstrate that the originally observed MIRI spectra were blueshifted by approximately 23\,nm at 5\,$\mathrm{\mu m}$ compared to the model spectra, with the best wavelength agreement occurring around 10\,$\mathrm{\mu m}$. Our results align more closely with the current official wavelength uncertainty provided for MIRI LRS spectra, which is accurate to within $\pm20$\,nm across most wavelengths, but slightly larger discrepancies exist at the 5\,$\mathrm{\mu m}$. This change is quite intuitive, as the JWST official documentation\footnote{\url{https://jwst-docs.stsci.edu/jwst-calibration-status/miri-calibration-status/miri-lrs-calibration-status\#gsc.tab=0}} states that, in December 2023 (with the 1174.pmap update), the wavelength calibration, as well as photometric correction and aperture correction of the MIRI LRS have been updated.

\subsection{The UltracoolSheet Catalog}

\begin{figure}[t]
\centering 
\includegraphics[width=0.45\textwidth]{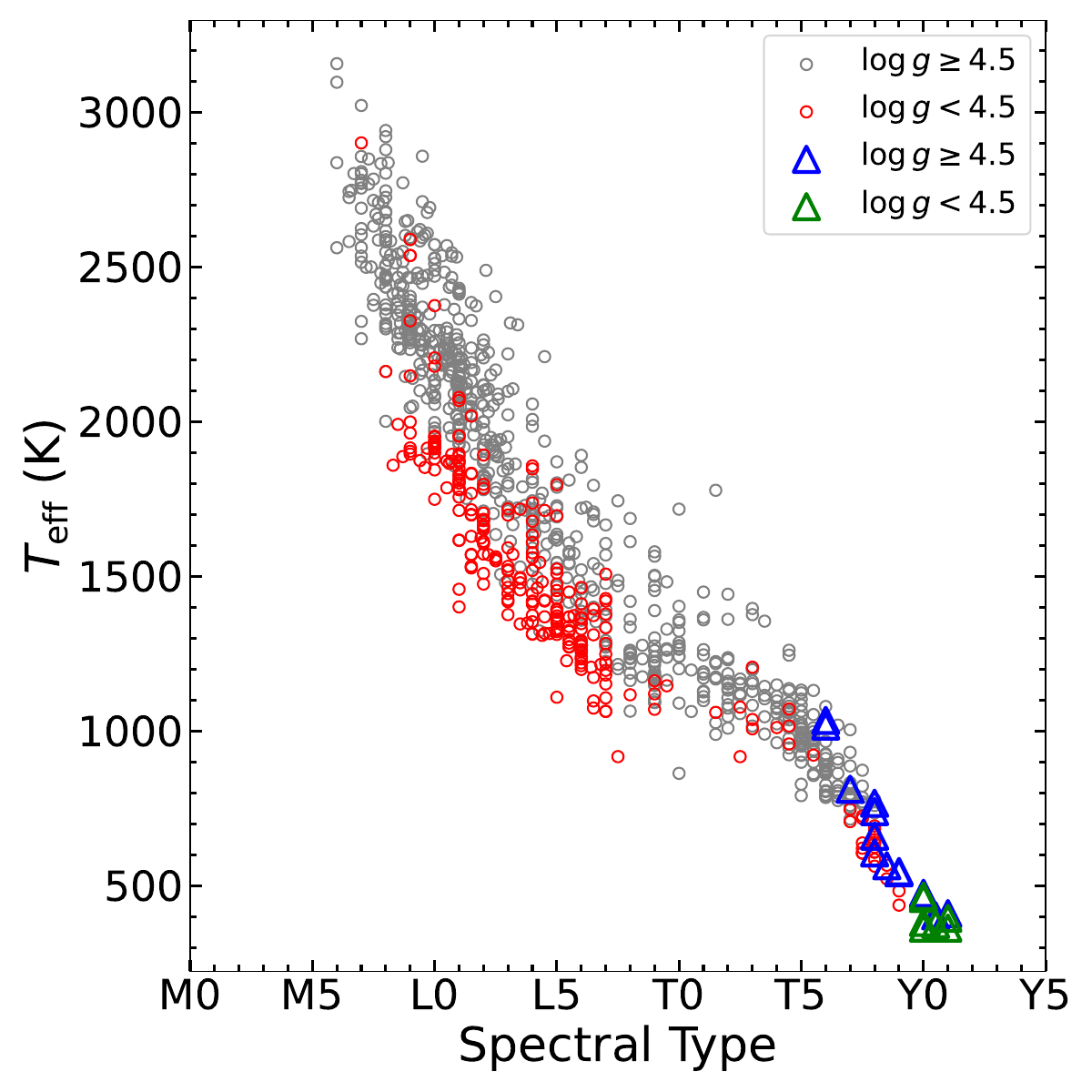}
\caption{Spectral type versus effective temperature for dwarfs. The data points represented by circles are sourced from The UltracoolSheet. The blue and green data points are our investigated brown dwarfs with physical parameters derived from the ATMO2020++ models. Different colored data points of the same shape represent data points in different ranges of surface gravity as shown in the legend.\label{fig:STvsTeff}} 
\end{figure}

The UltracoolSheet \citep{best_2024_10573247} is a comprehensive catalog that includes more than 4,000 ultracool dwarfs, primarily focusing on the spectral types M6 and beyond. Although it provides an extensive dataset, it contains relatively few late-type brown dwarfs due to the scarcity of detailed studies on these objects. 

In this section, we present a comparison between the dwarfs in The UltracoolSheet and our sample, which includes a more substantial collection of late-type brown dwarfs. Since our study focuses on comparing fundamental properties, we utilize the ``UltracoolSheet-FundamentalProperties'' table compiled by \citet{Sanghi2023ApJ}. This table contains over 1,000 ultracool dwarfs with both effective temperature and surface gravity derived from evolutionary models using the bolometric luminosities and age estimates (Detailed information and methodologies refer to Section 7 of \citealt{Sanghi2023ApJ}).

By comparing the key atmospheric parameters of our sample with those in The UltracoolSheet, we can thoroughly examine and complement the relationships between effective temperature, spectral type, and surface gravity, especially in the very low-temperature region.

\subsubsection{Spectral Type v.s. Effective Temperature}

The spectral type of a star typically correlates with its effective temperature. For T and Y dwarfs, effective temperatures generally fall below 1300\,K, while the spectral types of our sample range from T6 to $\geq$Y1, corresponding to even lower effective temperatures. There are relatively few studies on such cold (late-type) brown dwarfs.

The effective temperatures of the objects from The UltracoolSheet utilized here are derived from the ``teff\_evo'' parameter (evolutionary model-derived effective temperature). In Figure \ref{fig:STvsTeff}, circles represent data points from The UltracoolSheet, with grey circles indicating objects with $\log g\geq4.5$\,[$\mathrm{cm\,s^{-2}}$], and red circles indicating those with $\log g<4.5$\,[$\mathrm{cm\,s^{-2}}$]. For dwarfs with the same spectral type, those with lower surface gravity generally exhibit lower effective temperatures.

Our sample is concentrated in the spectral types $>$T5. The data points represented by triangles are derived from our sample, with effective temperatures and surface gravities obtained from the ATMO2020++ models. The corresponding spectral types are referenced in Table \ref{tab:samples}. The surface gravities derived from the Sonora Elf Owl models are mostly less than 4.5 and therefore are not discussed here.

The data points of our sample align well with the relationship between spectral type and effective temperature observed in The UltracoolSheet. The data points of our sample with $\log g\geq4.5$\,[$\mathrm{cm\,s^{-2}}$] are concentrated at spectral types $<$Y0. These data points, along with those from The UltracoolSheet, form well the relationship between $\log g$, $T_\mathrm{eff}$, and spectral type. On the other hand, the data points of our sample with $\log g<4.5$\,[$\mathrm{cm\,s^{-2}}$] are concentrated at spectral types $\geq$Y0, effectively complementing the sample in the very low effective temperature region.

\subsubsection{Surface Gravity v.s. Effective Temperature}

\begin{figure}[t]
\centering 
\hspace*{-1cm}\includegraphics[width=0.45\textwidth]{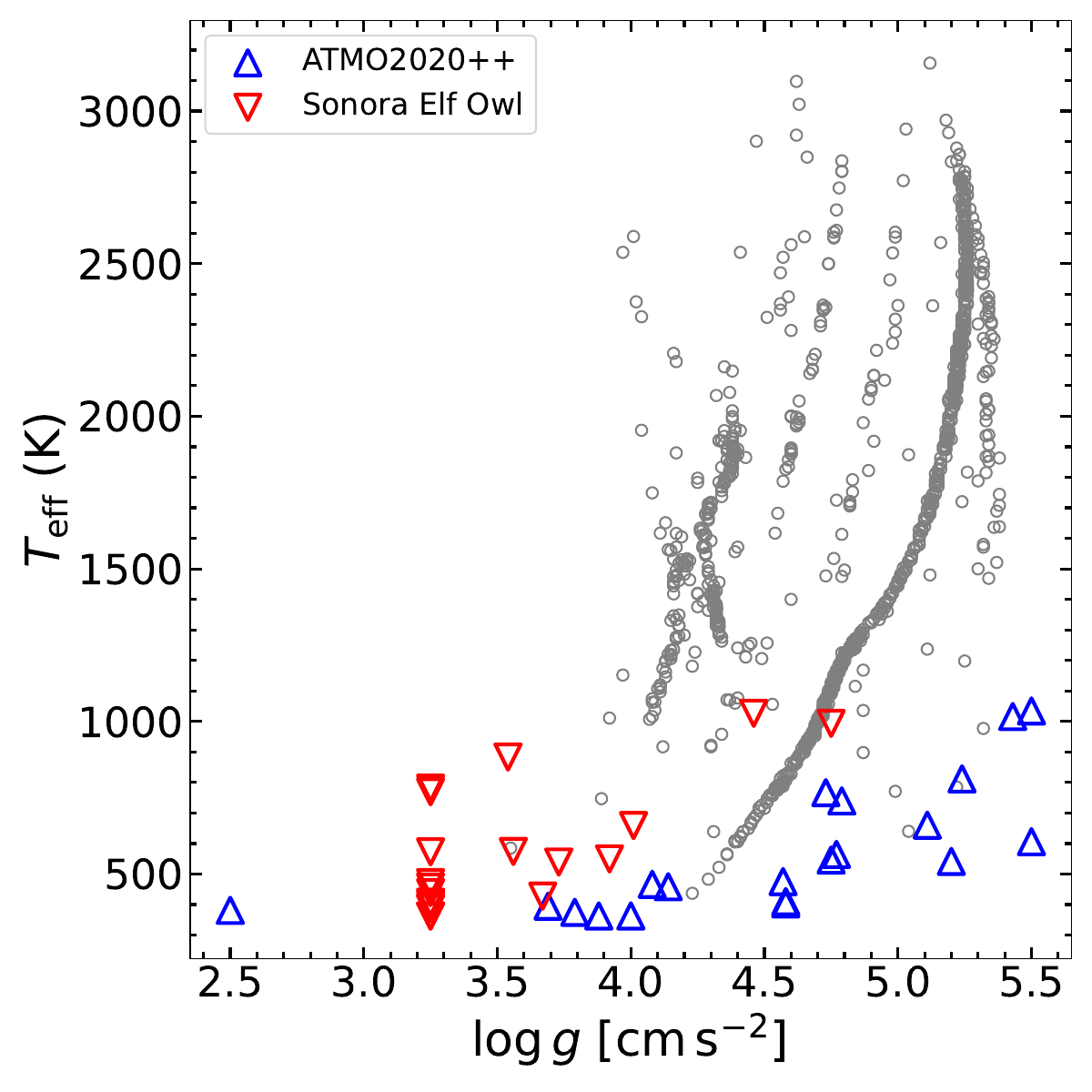}
\caption{Surface gravity versus effective temperature for dwarfs. These points in the grey circle are the objects from The UltracoolSheet. The points in the blue triangle are from our sample with the parameters derived using the ATMO2020++ models, while the red inverted triangles represent those parameters obtained using the Sonora Elf Owl models.\label{fig:loggvsTeff}}
\end{figure}

In Figure \ref{fig:loggvsTeff}, we examine the relationship between effective temperatures and surface gravities for our sample using the two different models, comparing them with the results of The UltracoolSheet. The values of $T_\mathrm{eff}$ and $\log g$ of the UltracoolSheet are derived from the evolutionary models. Grey circles represent data points from The UltracoolSheet, providing a broad overview of $\log g$ and $T_\mathrm{eff}$ distributions for dwarfs. Our sample data points are marked by specific symbols: blue triangles denote parameters derived using the ATMO2020++ models, while red inverted triangles indicate parameters obtained through the Sonora Elf Owl models.

A notable feature in the UltracoolSheet data points is the clustering observed in the $T_\mathrm{eff}$ range of 1000--2000\,K with $\log g$ values between 4 and 4.5. This cluster corresponds to objects of earlier spectral types where the $T_\mathrm{eff}$ values are constant. Additionally, there is a prominent band stretching from approximately 3000\,K at $\log g \approx 5.25$\,[$\mathrm{cm\,s^{-2}}$] down to around 500\,K at $\log g \approx 4.25$\,[$\mathrm{cm\,s^{-2}}$]. These different trends are primarily attributed to the varying methods of age estimation employed by \citet[][Section 6]{Sanghi2023ApJ}, as they utilized bolometric luminosities along with age estimates to derive surface gravity values through evolutionary models.

From the figure, it is evident that the parameters derived from our sample expand the existing data range, particularly at lower temperatures and surface gravities. Some data points, especially those where $\log g$ values converge to the model boundaries $\log g=3.25$ or 2.5\,[$\mathrm{cm\,s^{-2}}$], need to be analyzed with caution. Several data points represented by blue triangles can be connected to the end of the prominent band around 500\,K and $\log g \approx 4.25$\,[$\mathrm{cm\,s^{-2}}$]. Overall, most of the blue triangle data points are biased to be located on the right side of the prominent band, whereas the red inverted triangle data points are located almost exclusively on the left side.

\section{Conclusion} \label{sec:conclusion}

We have presented the spectra of 20 brown dwarfs, combining NIRSpec CLEAR/PRISM and MIRI LRS instruments on JWST, covering a wavelength range of 0.95--12\,$\mathrm{\mu m}$. Since the spectral types of our sample range from T6 to $\geq$Y1 and have relatively lower effective temperatures, they fill a gap in The UltracoolSheet catalog, complementing the relationships between the atmospheric parameters in the very low effective temperature region. The absolute flux calibration of the MIRI LRS spectra was corrected using the F1000W photometry with the official JWST pipeline and \textsc{DOLPHOT}. The results show that the uncertainty between the absolute flux calibration of the MIRI LRS and the photometry of the F1000W is around 4\%.

The Sonora Elf Owl and ATMO2020++ models, incorporating disequilibrium chemistry but different convective processes, were used to analyze their atmospheric parameters. Using the open-source code \textsc{UltraNest} with the nested sampling Monte Carlo algorithm “MLFriends”, we fitted the atmospheric parameters of our sample, as well as the wavelength correction parameters $A_\lambda$ and $C_\lambda$ of the MIRI LRS spectra and the scaling factor $\log (R^2/D^2)$ correlated with the radius and distance. The main results of this work are as follows.

\begin{enumerate}
    \item The model-fitting results showed highly consistent effective temperature between the two models, but deviations in other parameters. The surface gravities from the Sonora Elf Owl models are systematically lower than those from the ATMO2020++ models, on average by about 1 dex. Specifically, the $\log g$ for 12 out of 20 objects converges to a minimum value of $\log g=3.25$\,[$\mathrm{cm\,s^{-2}}$] using the Sonora Elf Owl models. The metallicities are essentially in the range of $-$0.3 to 0.4 using the Sonora Elf Owl models, and $-$0.3 to 0 using the ATMO2020++ models, both close to solar value. The additional introduction and utilization of the ATMO2020 models demonstrate that the reductions in the temperature gradient included in the ATMO2020++ models indeed affect the determination of surface gravity.
    \item Comparing the model spectra and observed spectra for the best-fit parameters of the two models revealed that deviations were mainly concentrated in the H$_2$O and CH$_4$ absorption features between 1 and 2.5\,$\mathrm{\mu m}$ and the CO$_2$ and CO molecular absorption bands between 4 and 5\,$\mathrm{\mu m}$. These regions are impacted by the interplay of nearly all atmospheric parameters and are also affected by model assumptions, such as the convective process.
    \item Through the examination of spectral sequences and the calculation of molecular spectral indices, we observed that the relative peak values around the H$_2$O and CH$_4$ absorption regions gradually decrease as the spectral type becomes later, while their spectral indices increase. The absorption features of CO$_2$ and CO molecules are more complex and vary between different sources, which may result from the different vertical mixing strengths and atmospheric metallicities. The spectral indices of CO$_2$ exhibit a slight decrease with later spectral types. Higher indices ($\mathrm{index}>1.4$) are more commonly observed in T dwarfs, whereas lower indices are prevalent among Y dwarfs. In contrast, the CO spectral indices in our sample show no significant trend with spectral type. For NH$_3$ indices, there is a general increase trend as the spectral type becomes later.
    \item By using the fitted parameters $\log(R^2/D^2)$ and the absolute parallaxes available in the literature, we obtain a radius range for our sample of approximately 0.8 to 1.2\,$R_{\mathrm{Jup}}$. We determine the masses and ages from the evolutionary tracks of \citet{Chabrier2023AA}, using the calculated radii and fitted effective temperatures from the two models. The masses span from 2 to 45\,$M_{\text{Jup}}$, with the majority below 30\,$M_{\text{Jup}}$. Furthermore, the age distribution of our sample ranges from 0.1 to 10\,Gyr, with most objects being younger than 6\,Gyr. For Y dwarfs, we found that their masses range between 2 and 20\,$M_{\mathrm{Jup}}$, and their ages span from 0.1 to 6.7\,Gyr.
    \item The results of atmospheric parameters using only NIRSpec CLEAR/PRISM or MIRI LRS spectra versus using the combined spectra show that the effective temperatures are generally consistent across the three cases (NIRSpec-only, MIRI-only, and combined). When only NIRSpec spectra are used, the $\log g$ values are similar to those obtained with combined spectra, but significant differences arise when only MIRI spectra are used. Metallicity is consistent with the combined spectra only when using the Sonora Elf Owl models and MIRI-only spectra. For the radii, the results derived from MIRI-only spectra demonstrate smaller deviations from the combined spectra, with differences generally within $\pm 0.1$\,$R_{\mathrm{Jup}}$. In contrast, the NIRSpec-only radii show larger discrepancies compared to those of combined spectra. Finally, the vertical eddy diffusion parameter $K_\mathrm{zz}$ and C/O ratio show significant discrepancies between values derived with NIRSpec-only or MIRI-only spectra and those obtained with combined spectra.
\end{enumerate}

Our work highlights the need for further refinement and validation of atmospheric models of brown dwarfs, particularly regarding surface gravity and metallicity. Observations and models must work hand in hand, and future JWST observations of brown dwarfs will be crucial for deepening our understanding of brown dwarf evolution and improving atmospheric models.

\section*{Acknowledgments} 
We sincerely thank the referee for the insightful and helpful suggestions and comments. We also would like to thank the author of the Sonora Elf Owl models, Sagnick Mukherjee, for the discussion of model-fitting results. This work is supported by the National Natural Science Foundation of China (NSFC) through projects 11988101, 12373028, 12133002, and 11933004. This work is also supported by the science research grants from the China Manned Space Project with No. CMS-CSST-220221-A09. S. W. acknowledges support from the Youth Innovation Promotion Association of the CAS with No. 2023065. Z. T. thanks Furen Deng for many useful discussions.
This work is based on observations made with the NASA/ ESA/CSA James Webb Space Telescope. The data were obtained from the Mikulski Archive for Space Telescopes (MAST) at the Space Telescope Science Institute. 

\software{NumPy \citep{numpy},
          SciPy \citep{scipy},
          Matplotlib \citep{matplotlib},
          UltraNest \citep{ultranest},  
          DOLPHOT \citep{dolphot1,dolphot2}, 
          corner.py \citep{corner}
          }

\appendix
\counterwithin{figure}{section}

\section{Test on the Reliability of Linear Interpolation} \label{app:Linear Interpolation Accuracy}

To assess the reliability of linear interpolation within the model grids, especially for the 5D model grids of the Sonora Elf Owl. we performed a series of leave-one-out validation tests using the Sonora Elf Owl model grids. Each test involved masking spectra for the fixed parameter value(s) and performing linear interpolation with the remaining model grids, subsequently comparing the interpolated results against the original model spectra. These interpolated spectra and original model spectra were convolved to the spectral resolutions of NIRSpec CLEAR/PRISM and MIRI LRS for better comparisons. The evaluation of these interpolations was quantified using the normalized root-mean-square error (NRMSE):
\begin{equation}
    NRMSE = \frac{\sqrt{\frac{1}{n}\sum_{i=1}^{n}(Y_i - \hat{Y}_i)^2}}{\hat{Y}_{\text{max}} - \hat{Y}_{\text{min}}},
\end{equation}
Where $Y$ is the linearly interpolated model spectrum, $\hat{Y}$ is the original model spectrum with the fixed parameters value(s). $\hat{Y}_{\text{max}}$ and $\hat{Y}_{\text{min}}$ are the corresponding maximum and minimum flux of the spectrum.

For instance, consider the parameter $\log K_{zz}$. We masked all spectra with $\log K_{zz} = 7$\,[$\mathrm{cm^2\,s^{-1}}$], including 5,760 model grids. The linear interpolation was then performed using the spectra with $\log K_{zz} \neq 7$\,[$\mathrm{cm^2\,s^{-1}}$]. The linearly interpolated results at $\log K_{zz} = 7$\,[$\mathrm{cm^2\,s^{-1}}$] were compared to the original model spectra, yielding an average NRMSE of 0.006 with 5,760 spectra.

Similar tests were conducted for other parameters as detailed below:

\begin{itemize}
    \item ($T_{\text{eff}}$): We masked the spectra with $T_{\text{eff}} = 400$\,K, 600\,K, and 1000\,K, including 1,200 model grids for each $T_{\text{eff}}$. The average NRMSE for these $T_{\text{eff}}$, in ascending order, are 0.014, 0.007, and 0.006, respectively.
    \item ($\log g$): Tests on $\log g = 3.75$\,[$\mathrm{cm\,s^{-2}}$] and 4.75\,[$\mathrm{cm\,s^{-2}}$] including 2880 model grids for each $\log g$, yielded average NRMSE of 0.008 and 0.006 for the two $\log g$ values.
    \item ($[\mathrm{M/H}]$): The value $[\mathrm{M/H}] = 0$ was tested. The mean NRMSE is 0.016 with 4800 model grids.
    \item  (C/O): Spectra for C/O$ = 0.458$ including 7200 model grids were removed. The resulting mean NRMSE is 0.008.
\end{itemize}
Across all parameters, the linearly interpolated spectra closely matched the masked spectra in most wavelength regions, with small NRMSE values. 

\begin{figure*}
\centering 
\includegraphics[width=1.\textwidth]{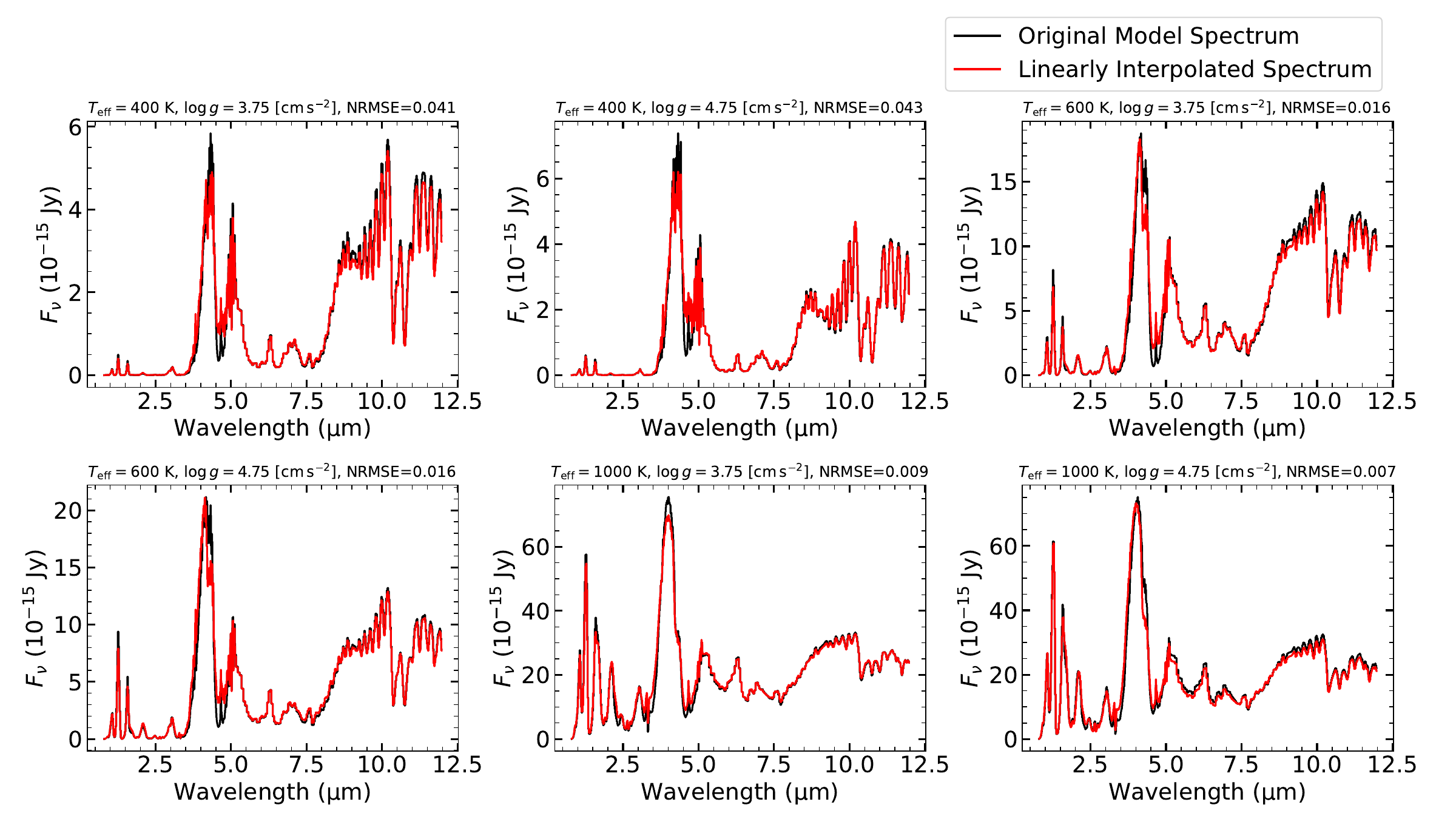}
\caption{Comparisons between the original model spectra and the linearly interpolated spectra with the same parameters using the Sonora Elf Owl models. All six samples have $\log K_{zz} = 7$\,[$\mathrm{cm^2\,s^{-1}}$], $[\mathrm{M/H}] = 0$, and C/O $= 0.458$. Other parameters are provided at the top of each subfigure. \label{appfig:interpolate}} 
\end{figure*}

We further tested the reliability of linear interpolation by simultaneously removing the parameter values specified above from each parameter space and linearly interpolating with the remaining parameters across all five dimensions. Specifically, we removed three values for $T_\mathrm{eff}$, two values for $\log g$, and one value for each of the other parameters mentioned above. This provided us with a total of six samples for comparison. We then compared the resulting five-dimensional linear interpolation spectra with the original model spectra as shown in Figure~\ref{appfig:interpolate}.

From Figure~\ref{appfig:interpolate}, the variations in $\log g$ result in relatively minor changes in NRMSE values. In contrast, changes in $T_\mathrm{eff}$ significantly impact the interpolation errors, with the largest NRMSE observed at 400\,K. As $T_\mathrm{eff}$ increases, the NRMSE decreases, indicating a trend of smaller NRMSE values at higher $T_\mathrm{eff}$. This is consistent with the trend of the NRMSE values obtained from the above test with only $T_\mathrm{eff}$.

Comparing the interpolation errors in different wavelength regions, it is clear that the linearly interpolated spectra are well-matched to the model spectra in most wavelength regions, but a relatively large discrepancy occurs in the wavelength region 4--5\,$\mathrm{\mu m}$. These noticeable discrepancies are likely due to the strong and overlapping molecular absorption features, particularly from CO$_2$, CO, CH$_4$, and H$_2$O, leading to this region vary with not only $T_\mathrm{eff}$ and $\log g$, but also [M/H] and $\log K_{zz}$ (See Section~\ref{subsec:Partial_spec}). The dependence of multiple parameters in this wavelength region could likely cause non-linear flux changes. 

\begin{deluxetable}{ccccccccc}[t]
\tablecaption{Comparisons between model parameters and fitted parameters\label{tab:inter_test}}
\setlength{\tabcolsep}{0.45 cm}
\tablehead{\colhead{} & \colhead{$\log K_\mathrm{zz}$} & \colhead{$T_\mathrm{eff}$} & \colhead{$\log g$} & \colhead{[M/H]} & \colhead{C/O}& \colhead{$A_\lambda$}& \colhead{$C_\lambda$}& \colhead{$\log (R^2/D^2)$}\\ 
\colhead{} & \colhead{$[\mathrm{cm^2\,s^{-1}}]$} & \colhead{(K)} & \colhead{$[\mathrm{cm\,s^{-2}}]$} & \colhead{} & \colhead{}& \colhead{($\times 10^{-3}$)}& \colhead{($\times 10^{-3}\,\mu$m)}& \colhead{}} 
\startdata
\multirow{2}{*}{model\_1} & 7    & 400    & 3.75 & 0     & 0.458 & 0 & 0 & -20 \\
&8.34& 383.44& 3.63& -0.04& 0.625& -0.62& 5.65& -19.93 \\ \hline
\multirow{2}{*}{model\_2} & 7    & 400    & 4.75 & 0     & 0.458 & 0 & 0 & -20 \\
&9.00& 385.83& 4.69& -0.02& 0.605& -0.34& 2.97& -19.97 \\ \hline
\multirow{2}{*}{model\_3} & 7    & 600    & 3.75 & 0     & 0.458 & 0 & 0 & -20 \\
&8.65& 579.25& 3.75& -0.02& 0.640& -0.49& 3.91& -19.94  \\ \hline
\multirow{2}{*}{model\_4} & 7    & 600    & 4.75 & 0     & 0.458 & 0 & 0 & -20 \\
&8.91& 577.24& 4.68& -0.02& 0.643& -0.49 & 3.80& -19.95 \\ \hline
\multirow{2}{*}{model\_5} & 7    & 1000   & 3.75 & 0     & 0.458 & 0 & 0 & -20 \\
&7.93& 960.22& 3.88& 0.04& 0.566& -0.34& 2.93& -19.94 \\ \hline
\multirow{2}{*}{model\_6} & 7    & 1000   & 4.75 & 0     & 0.458 & 0 & 0 & -20 \\
&8.50& 969.29& 5.03& 0.00& 0.658& -0.36& 2.89& -19.95 \\ \hline
\enddata
\tablecomments{The first row of each model is the model parameter values and the second row is the fitted values.}
\end{deluxetable}

Subsequently, we tested the impact of linear interpolation errors on the atmospheric parameters by using the six model spectra mentioned above as mock observed spectra and fitting them with the nested sampling method outlined in Section~\ref{subsec:method}. The linearly interpolated model spectra were generated from model spectra for all other parameter values, masking those corresponding to the six specified models. The wavelength correction parameters for these model spectra were set to zero, and the model spectra were scaled by a factor corresponding to $\log(R^2/D^2) = -20$. The flux errors were set to 1 per cent of the fluxes. 

The fitting results are presented in Table~\ref{tab:inter_test}, where the first row for each model lists the original model parameters, and the second row shows the fitted values. The fitting errors are not presented because the flux errors were set arbitrarily. We calculated the mean and standard deviation of the differences between the fitted and model values for each parameter: $1.56 \pm 0.36$\,[$\mathrm{cm^2\,s^{-1}}$] for $\log K_{zz}$, $-24.12 \pm 8.73$\,K for $T_{\text{eff}}$, $0.03 \pm 0.14$\,[$\mathrm{cm\,s^{-2}}$] for $\log g$, $-0.01 \pm 0.03$ for [M/H], $0.16 \pm 0.03$ for C/O, $(-0.44 \pm 0.10) \times 10^{-3}$ for $A_\lambda$, $(3.69 \pm 0.96) \times 10^{-3}\,\mu\mathrm{m}$ for $C_\lambda$, and $0.05 \pm 0.01$ for $\log(R^2/D^2)$. 

Based on these results, the interpolation errors have a small impact on [M/H] and the wavelength correction parameters. For $ T_{\text{eff}} $ and $\log(R^2/D^2)$, these two parameters exhibit degeneracy and tend to be negatively correlated within certain ranges, where a smaller $\log(R^2/D^2)$ is often associated with a higher $ T_{\text{eff}} $. Therefore, their uncertainties are partially influenced by each other’s fits. The linear interpolation error, on the other hand, significantly impact $\log K_{zz}$ and C/O, indicating that caution is needed when considering the two parameters.

Our findings are generally consistent with the results of \citet{Zhang2021ApJ}, who utilized a Starfish-based Bayesian approach (incorporating interpolation errors) with the Sonora Bobcat models, yielding parameter uncertainties of about 28\,K for $T_{\text{eff}}$, 0.29\,[$\mathrm{cm\,s^{-2}}$] for $\log g$, and 0.15 for [M/H]. They also compared the fitting results between the traditional Bayesian approach (only including the observed flux errors) and the Starfish-based Bayesian approach. The results indicate that the parameter estimates from both methods are generally consistent, but the Starfish-based Bayesian method typically provides uncertainties that are larger by a factor of 10 compared to the traditional Bayesian approach.

\section{Corner plots of the sample} \label{app:corner}
\begin{figure*}
\centering 
\includegraphics[width=1.\textwidth]{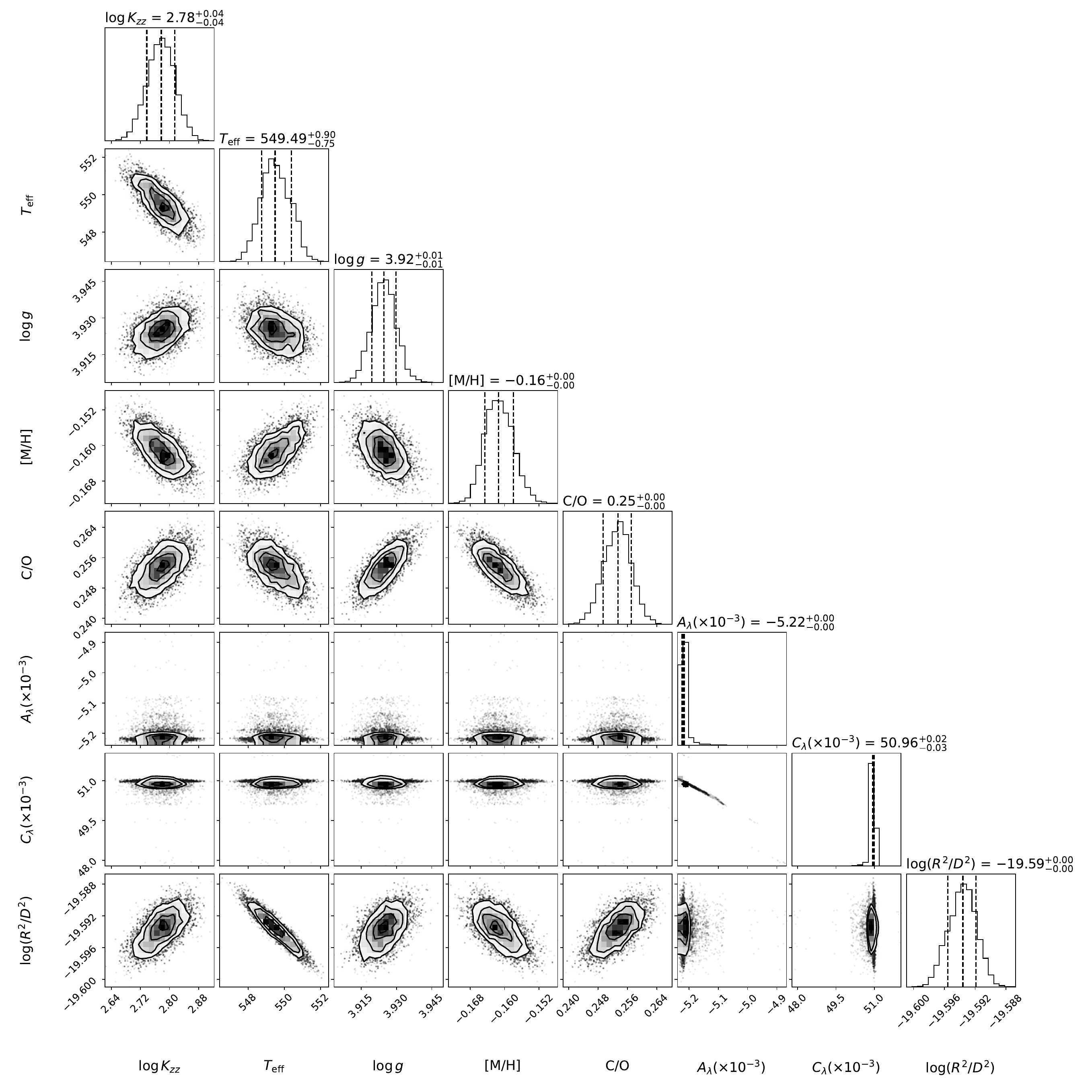}
\caption{The marginalized posterior distribution of eight parameters for the observed spectra of WISEA J2159$-$48, using the Sonora Elf Owl models. The parameters are, in order, the logarithm of vertical eddy diffusion parameter $\log K_\mathrm{zz}$, effective temperature $T_\mathrm{eff}$, surface gravity $\log g$, metallicity [M/H], carbon–to–oxygen ratio C/O, wavelength correction parameters $A_\lambda$ and $C_\lambda$, and scaling factor $\log(R^2/D^2)$. \label{fig:corner_owl_2159}} 
\end{figure*}

\begin{figure*}
\centering 
\includegraphics[width=1.\textwidth]{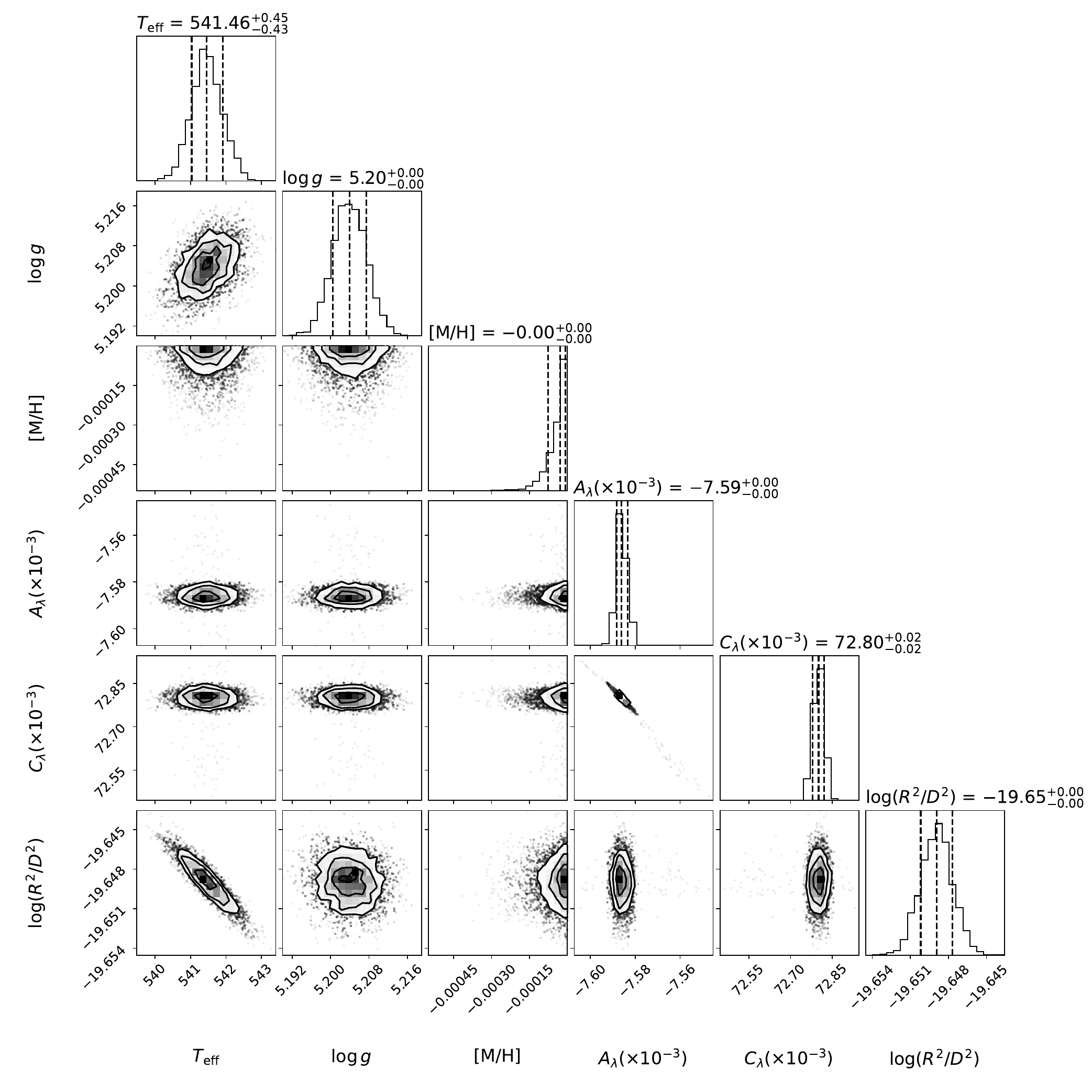}
\caption{The marginalized posterior distribution of six parameters for the observed spectra of WISEA J2159$-$48, using the ATMO2020++ models. The parameters are, in order, effective temperature $T_\mathrm{eff}$, surface gravity $\log g$, metallicity [M/H], wavelength correction parameters $A_\lambda$ and $C_\lambda$, and scaling factor $\log(R^2/D^2)$.\label{fig:corner_atmo_2159}} 
\end{figure*}

\begin{figure*}
\centering 
\includegraphics[width=1.\textwidth]{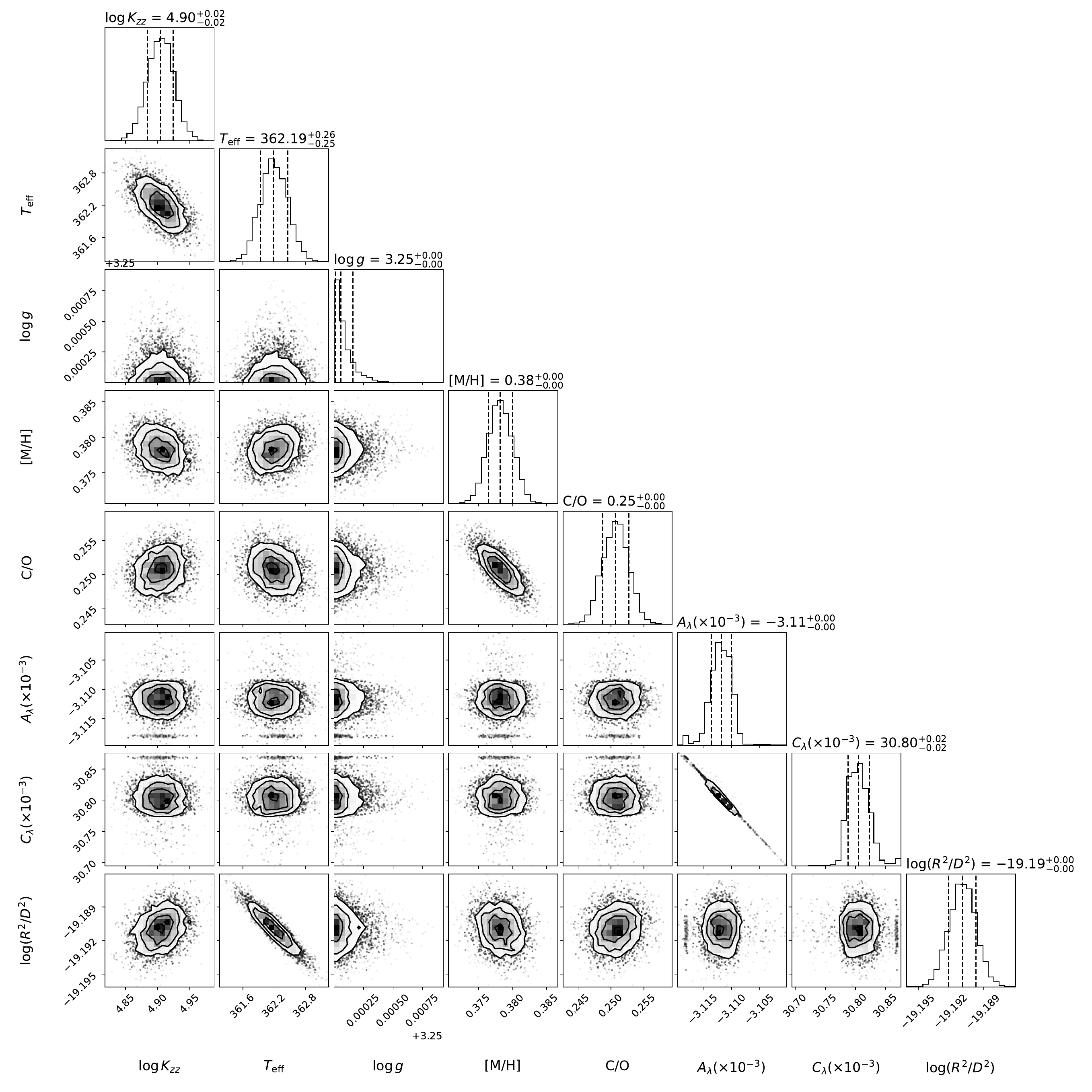}
\caption{Similar to Figure~\ref{fig:corner_owl_2159}, but for CWISEP J1446$-$23 using the Sonora Elf Owl models.\label{fig:corner_owl_1446}} 
\end{figure*}

\begin{figure*}
\centering 
\includegraphics[width=1.\textwidth]{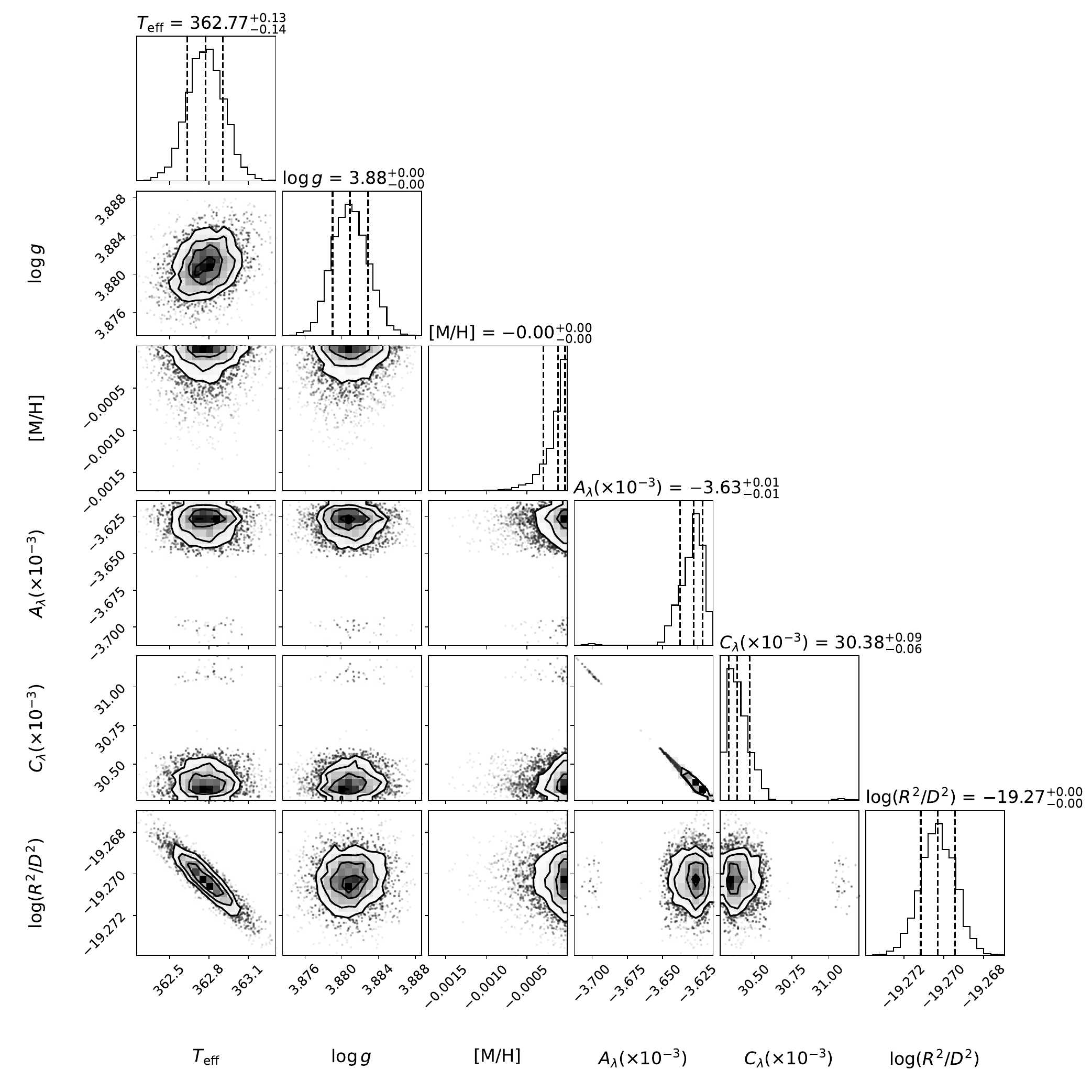}
\caption{Similar to Figure~\ref{fig:corner_atmo_2159}, but for CWISEP J1446$-$23 using the ATMO2020++ models.\label{fig:corner_atmo_1446}} 
\end{figure*}

Figure~\ref{fig:corner_owl_2159} and Figure~\ref{fig:corner_atmo_2159} present the corner plot for WISEA J2159$-$48, derived using the Sonora Elf Owl and ATMO2020++ models, respectively. Similarly, Figure~\ref{fig:corner_owl_1446} and Figure~\ref{fig:corner_atmo_1446} show the corresponding plots for CWISEP J1446$-$23. We note that the step-like behavior of the $A_\lambda$ and $C_\lambda$ could be attributed to the limited wavelength resolution of the model spectra and the small uncertainties of the observed spectra. The complete set of corner plots for all sources in our sample is available online\footnote{\url{https://github.com/LuShenJ/Parameters_of_20_Cold_Brown_Dwarfs_in_JWST}}.

\bibliography{main}{}
\bibliographystyle{aasjournal}

\end{document}